%% file: main.tex
\pdfoutput=1
\documentclass[11pt]{article}

\usepackage[margin=1in]{geometry}
\usepackage{indentfirst}
\usepackage{amsmath,amssymb}
\usepackage{bbold}
\usepackage{tikz}
\usepackage{graphicx}
\usepackage{subcaption}
\usepackage{booktabs}
\usepackage{authblk}
\usepackage[section]{placeins}
\usepackage[hidelinks]{hyperref}

\usetikzlibrary{calc, shapes.misc}

\title{Attacking P2P Networks:\\ Liveness Attacks on the GossipSub and DOG Gossip Protocols}

\author[1]{Jiaheng Li}
\author[1]{Anton Paramonov}
\author[2]{Yann Vonlanthen}
\affil[1]{ETH Zurich}
\affil[2]{Ethereum Foundation}
\date{}

\begin{document}

\maketitle

\begin{abstract}
  Peer-to-peer (P2P) networks form the communication layer of large decentralized
  systems such as Ethereum, and the correctness of higher-level protocols depends on the
  gossip layer delivering messages to honest nodes. We study the robustness of two gossip
  protocols against adversaries whose only goal is to violate liveness, that is, to prevent
  honest nodes from receiving fresh messages, without forging, altering, or invalidating any
  message.

  For GossipSub, the publish-subscribe protocol used by Ethereum, we show that an
  adversary operating a constant number of geographically distributed nodes with a fixed
  latency advantage can take over the ``First Message Deliveries'' peer-scoring term and
  separate a constant fraction of nodes from the honest mesh. We analyze the attack under
  a geographical model, address the random-peer and outbound-connection mechanisms, and
  show that the separation can be maintained under realistic churn.

  For DOG, a transaction-dissemination protocol designed for CometBFT, we first give
  simple counterexamples in which route filtering disconnects the dissemination graph even
  without failures or Byzantine participants, and then present a practical attack that uses
  only a single physical node with a constant number of identities and legitimate actions to
  disable every forwarding route, separating the full network while allowing only direct
  publisher-to-neighbor delivery. We verify the attack in a simulated network and analyze
  its cost, assumptions, and possible mitigations.
\end{abstract}

\input{sections/intro.tex}
\input{sections/gossipsub.tex}
\input{sections/dog.tex}

\input{sections/bibliography.tex}

\end{document}

%% file: sections/intro.tex

\section{Introduction}
\label{sec:intro}

\subsection{Background}

Peer-to-peer (P2P) networks are distributed systems in which participating nodes
communicate directly and collectively provide a service without relying on a single central
coordinator. In large decentralized systems such as Ethereum, the P2P network forms the
communication layer over which nodes disseminate transactions, blocks, and
protocol-control messages. The correctness of higher-level mechanisms, including
consensus, therefore depends not only on their own algorithms but also on the timely and
reliable delivery of messages by the underlying P2P network.

Gossiping is a family of dissemination techniques in which each node forwards a newly
received message to a selection of its peers, causing the message to propagate through the
network by repeated local exchanges. In blockchain systems, gossip protocols are used to
disseminate transactions, blocks, and consensus messages among nodes. For example,
Ethereum relies on gossiping to propagate the information that validators and other nodes
require to maintain a consistent view of the system and participate in consensus.

Although consensus protocols have received extensive formal analysis, the security
properties of the P2P protocols that carry their messages are often understood less
rigorously. This disparity is significant because a secure consensus protocol cannot
compensate for a communication layer that prevents honest nodes from receiving the
messages required to participate. An adversarial analysis of gossip protocols is therefore
necessary to identify whether their optimization mechanisms introduce exploitable failure
modes that can delay or suppress message delivery and thereby threaten the availability of
the systems built upon them.

\subsubsection{Gossiping Algorithms}

Flooding is the simplest gossiping strategy: whenever a node first receives a message, it
forwards that message to every eligible peer. This strategy is robust when the network is
connected, because it explores every available propagation edge, but it produces
substantial redundant traffic and consumes bandwidth and processing capacity at every
node.

Practical gossip protocols therefore try to reduce this overhead. For example, they may
select only some peers to forward messages to, suppress propagation along routes judged
redundant, or construct a smaller dissemination mesh.

In this report, a network (or a mesh) is modeled as a graph whose vertices are
participating nodes and whose edges represent peer relationships over which gossip messages
may be exchanged. Two nodes connected by such an edge are called peers, and the peers
adjacent to a node constitute its local view of the network.

\subsection{Goal}

The objective of the attacks studied in this report is to violate the liveness of the gossip
layer. We say that the gossip layer satisfies liveness for a message if every continuously
participating honest node that is intended to receive the message eventually receives it.
Accordingly, an attack succeeds against a target node when it can indefinitely prevent that
node from receiving fresh messages disseminated by honest nodes, and it succeeds against
the full network when it can partition the effective dissemination graph so that honest
messages cannot propagate between its components. This objective does not require the
adversary to forge, alter, or invalidate messages; preventing their delivery is sufficient.

Such a failure may delay transaction inclusion, leave nodes with inconsistent views of
the system, and obstruct the progress of higher-level protocols that depend on
communication among honest participants.

\subsection{General Methods}

The general attack strategy is to exploit the protocol's own peer- or route-selection
mechanism until every effective incoming propagation path of a target node is controlled by
the adversary. The adversary first causes honest alternatives to be filtered, pruned, or
scored below attacker-controlled alternatives while continuing to behave sufficiently well
for its own connections to be retained, if necessary. Once the attacker controls or disables
all selected propagation paths to the target, it may withhold subsequent messages and the
target will be separated from the honest dissemination network. Repeating this procedure
for multiple targets can remove enough effective edges to partition a large portion of the
mesh, even when the underlying transport connections have not all failed or been removed.

\subsection{Possible Capabilities of Attackers}
\label{sec:capabilities}

Depending on the attack under consideration, we may grant the adversary some or all of
the following capabilities:

\begin{enumerate}
  \item operating either a limited number of attacker-controlled nodes or a large number of
        such nodes at different geographical locations;
  \item holding and using either a limited number of network identities or a large number of
        such identities, such as distinct IP addresses, which may be used by the same physical
        attacker node;
  \item controlling the behavior of attacker nodes, including which peers they connect to and
        which messages they forward, withhold, delay, or send to selected recipients;
  \item publishing (limited or unlimited) fresh valid application messages; and
  \item providing lower-latency communication between attacker nodes than is typical between
        honest nodes.
\end{enumerate}

The first capability reflects the permissionless nature of the P2P networks considered in
this report, in which an ordinary participant may join the network and establish connections
without compromising an existing honest node. The number and geographical distribution of
attacker nodes are treated as resources available to the adversary.

The second capability reflects that an adversary may hold and use multiple identities,
such as IP addresses, without incurring the cost of deploying them on separate devices. This
capability is relevant when the attacker needs to make multiple distinct connections to a
target.

The third capability follows from the adversary operating its own protocol implementations
and therefore choosing how its own nodes connect and forward messages. It permits
deviations from the prescribed protocol by attacker nodes, but it does not permit the
adversary to alter messages exchanged between honest nodes or to dictate the behavior of
honest nodes. Such deviations may be local and may not be easily recognized by honest nodes
as malicious.

The fourth capability is considered because attacker nodes generally have the same right
as honest nodes to originate valid application messages. However, although the P2P protocol
may allow any participant to publish messages, the higher-level application may restrict who
can produce a valid message or how frequently such messages can be produced. For example,
producing a valid vote in Ethereum requires control of a validator with the corresponding
voting power and is subject to the rules of the consensus protocol. We therefore consider
separately whether an attacker can publish fresh valid messages, and whether the frequency
is limited.

The fifth capability models an adversary that invests in better network infrastructure,
for example by connecting its geographically distributed nodes through a high-performance
private network. This latency advantage is relevant because message arrival order can affect
how a gossip protocol selects propagation routes or peers.

\subsection{Results Summary}

We study two gossip protocols: GossipSub~\cite{gossipsub-spec}, which is used by Ethereum,
and DOG~\cite{informal-dog, cometbft-dog-spec}, which is designed for use by
CometBFT~\cite{cometbft}.

\subsubsection{Attacking GossipSub}

An unlimited supply of fresh messages under Capability~4 permits a comparatively direct
attack on GossipSub, but we do not focus on this case because the corresponding
application-level assumption may be impractical.

Instead, our main attack uses a constant number of geographically distributed physical
nodes under Capability~1 and a larger, but still constant, number of network identities
under Capability~2. It uses Capability~3 to control the connections and message forwarding
of the attacker nodes. It also assumes Capability~5 in the form of a fixed multiplicative
latency advantage for communication between attacker nodes relative to typical honest
links.

In addition, for the analysis, we approximate the geographical distribution of ordinary
nodes by sampling them uniformly from a spatial region, and we additionally consider a
constant number of highly concentrated node clusters. We also assume that each fresh message
originates from a node selected uniformly from the network. A more formal model is
introduced in Section~\ref{sec:gossipsub-model}.

Under the geographical model analyzed in this report, these capabilities allow the
attacker to separate a constant fraction of nodes from the ordinary GossipSub mesh, before
accounting for the outbound-connections quota.

The outbound-connections quota requires separate treatment because it preserves peers
selected through outbound connections even when the attack controls the other peer-selection
mechanisms. To attack these connections, the adversary must also hold a small fraction of
the identities in the candidate pool from which a target samples a replacement peer. This
argument also assumes that the candidate pool contains all network nodes or is sampled
uniformly from them. For example, in the model with 100 geographically distributed attacker
nodes and private-network latency equal to 80\% of typical honest-link latency, the analysis
considers attacker identities comprising 1\%--2\% of the candidate pool.

Finally, our analysis considers message delivery through mesh peers but does not account
for delivery from non-mesh peers through GossipSub's metadata layer using \texttt{IHAVE} and
\texttt{IWANT} messages. Separation from the mesh does not by itself establish that a target
receives no messages through these metadata mechanisms. Because the metadata mechanisms are
largely independent of the mesh, our method does not apply to that layer. The mechanisms
that form the metadata-exchange network are also specified largely outside the GossipSub
protocol, so they are not analyzed in this study.

\subsubsection{Attacking DOG}

The DOG attack requires only one physical attacker node and therefore imposes no
requirement on its geographical location under Capability~1. Under Capability~2, this node
uses a constant number of network identities, with five logical attacker nodes (network
identities) being sufficient under the default configuration in the construction presented in
this report. The attack also uses Capability~3 to control the forwarding behavior and timing
of these logical nodes. For simplicity, our principal construction grants the attacker
Capability~4 with an unlimited frequency of fresh messages. The attack can instead proceed
gradually when fresh messages are available only at a limited frequency, and messages
published by honest nodes may also be used in place of attacker-originated messages in some
stages. Consequently, unlimited publication is a simplifying assumption rather than an
essential requirement of the attack.

The attack can separate the full network by disabling every forwarding route, while
messages may still travel directly from their publishers to adjacent peers over mesh links.
Thus, a newly published message may reach the publisher's direct neighbors, but it cannot be
forwarded beyond them. The attacker can maintain this separation indefinitely by preventing
DOG from resetting the disabled routes. The cost of establishing the separation is
proportional to the number of two-hop routes, each consisting of a source, one intermediate
forwarder, and a target, which is the basic unit filtered by DOG. The cost of maintaining the
separation is proportional to the number of fresh messages subsequently published in the
network, with additional forwarding work determined by the number of direct neighbors of
each publisher.

\begin{table}[htb]
  \centering
  \caption{Summary of the capabilities, results, and limitations of the two attacks.}
  \label{tab:summary}
  \small
  \begin{tabular}{@{}p{0.11\textwidth}p{0.40\textwidth}p{0.40\textwidth}@{}}
    \toprule
                     & \textbf{GossipSub}
                     & \textbf{DOG} \\
    \midrule
    Capability 1     & A constant number of geographically distributed physical nodes.
                     & One physical node with no geographical-distribution requirement. \\[2pt]
    Capability 2     & A larger but still constant number of identities, with additional
                       presence in the replacement-peer pool required to attack outbound
                       connections.
                     & A constant number of identities, with five logical nodes (network
                       identities) sufficing under the default configuration in our
                       construction. \\[2pt]
    Capability 3     & Control over peer connections and message forwarding.
                     & Control over forwarding, recipients, and message timing. \\[2pt]
    Capability 4     & Not required by the main attack, although unlimited publication enables
                       a simpler attack that we do not focus on.
                     & Unlimited fresh messages in the principal construction, while
                       limited-frequency or honest-originated messages can also be used. \\[2pt]
    Capability 5     & A fixed multiplicative latency advantage between attacker nodes.
                     & Not required. \\[2pt]
    Additional model & Ordinary nodes are distributed uniformly over a spatial region, with a
                       constant number of additional highly concentrated node clusters. Fresh
                       messages originate uniformly among the network nodes.
                     & \\[2pt]
    Result           & A constant fraction of nodes is separated from the ordinary mesh under
                       the analyzed geographical model.
                     & All forwarding routes are disabled and full separation can be maintained
                       indefinitely, while direct delivery from a publisher to its neighbors
                       remains possible. \\[2pt]
    Limitations      & The outbound-quota argument assumes uniform peer sampling and sufficient
                       attacker identities in the candidate pool. Non-mesh delivery through
                       \texttt{IHAVE}/\texttt{IWANT} is not analyzed.
                     & \\
    \bottomrule
  \end{tabular}
\end{table}

%% file: sections/gossipsub.tex

\section{Attacking GossipSub}
\label{sec:gossipsub}

\subsection{The GossipSub Protocol}

GossipSub~\cite{gossipsub-spec} is a topic-based publish-subscribe protocol for
disseminating messages in a P2P network. A topic is an application-defined channel that
identifies a class of messages, and nodes may subscribe to the topics whose messages they
wish to receive. For each topic, participating nodes form a mesh in which every node selects
a subset of the other participants as its mesh peers. When a node receives a new valid
message from a mesh peer, it forwards the full message eagerly to its other mesh peers,
allowing the message to propagate through the mesh. GossipSub limits the number of mesh
peers selected by each node to reduce the redundant traffic produced by flooding, while
retaining additional known peers for metadata-based dissemination and mesh maintenance.

\subsubsection{Mesh, Grafting and Pruning}

A node adds a mesh peer by sending it a \texttt{GRAFT} control message, and it removes a
mesh peer by sending it a \texttt{PRUNE} control message. When a node joins a topic, it
selects and grafts peers until it has the target number $D$ of mesh peers. During periodic
mesh maintenance, if the node has fewer than $D_\text{lo}$ mesh peers, it grafts eligible
peers until their number returns toward $D$. If the node has more than $D_\text{hi}$ mesh
peers, it selects peers to retain and prunes the remaining peers until their number returns
toward $D$.

When a node needs to graft some peers, the graft targets are selected from the node's
locally known eligible participants in the topic, while the mechanism by which these peers
are discovered is not specified by GossipSub itself. This locally known set of topic
participants is the same set that we later define as the node's meta peers when discussing
metadata exchange during heartbeats.

Throughout this chapter, we use the default mesh configuration of the Prysm~\cite{prysm}
implementation of Ethereum unless stated otherwise. This configuration sets $D = 8$,
$D_\text{lo} = 6$, and $D_\text{hi} = 12$.

\subsubsection{Peer Selection}

When a node has too many mesh peers, it retains the $D_\text{score}$ mesh peers with the
highest peer scores~\cite{gossipsub-v11} and selects the remaining peers to retain randomly.
The selection is additionally constrained by an outbound-connections quota, which requires
the retained set to contain at least $D_\text{out}$ peers connected through outbound
connections.

Combining score-based and random selection preserves well-performing peers without making
the retained peer set entirely deterministic or preventing new peers from participating.

Under the default configuration, $D_\text{score} = 4$ and $D_\text{out} = 2$.

\paragraph{Outbound Connections Quota.}
An outbound connection of a node is a transport connection initiated by that node,
independently of the direction in which application messages later travel over the
connection. When a node has too many mesh peers, it preserves at least $D_\text{out}$
outbound peers by replacing lower-priority retained peers if necessary. During heartbeat
maintenance, if the node already has at least $D_\text{lo}$ mesh peers but fewer than
$D_\text{out}$ outbound mesh peers, it grafts enough eligible outbound peers to satisfy the
quota.

The outbound quota prevents an adversary that creates only inbound connections to a target
from occupying all of the target's mesh-peer positions.

\paragraph{Peer Scoring.}
Each node locally computes a score for every known peer from its own observations of that
peer's behavior, and these scores are not shared between nodes. Peer scores influence which
peers are retained during pruning, which peers are eligible for grafting, and whether control
messages from a peer are accepted. The score is a weighted combination of topic-specific and
global components that reward useful delivery behavior and penalize unreliability or
misconduct.

The score contains the following topic-specific components~\cite{gossipsub-v11}:
\begin{itemize}
  \item \emph{Time in Mesh} gives a bounded positive contribution according to the time for
        which the peer has remained a mesh peer.
  \item \emph{First Message Deliveries} gives a positive contribution when the peer is the
        first to deliver a new valid message, up to a configured cap.
  \item \emph{Mesh Message Deliveries} penalizes a mesh peer if its rate of first or
        near-first deliveries remains below a configured threshold after an activation period.
  \item \emph{Mesh Message Delivery Failures} records the delivery deficit of a peer when it
        is pruned and retains this penalty after the peer leaves the mesh.
  \item \emph{Invalid Messages} penalizes a peer for messages that fail application-specific
        validation.
\end{itemize}

The score also contains the following global components:
\begin{itemize}
  \item The \emph{Application-Specific Score} allows the application to assign an additional
        score according to its own policy.
  \item The \emph{IP Colocation Factor} penalizes peers when too many known peers use the
        same IP address.
  \item The \emph{Behavioral Penalty} penalizes protocol violations such as attempting to
        graft before the applicable backoff period has expired.
\end{itemize}

Among these components, First Message Deliveries is particularly important to normal
message propagation. It rewards peers that deliver new valid messages before competing peers.
Time in Mesh provides a bounded positive contribution that favors peers with sustained mesh
participation. Mesh Message Deliveries is instead a penalty component: it contributes no
penalty when delivery is at or above the configured threshold and contributes a deficit-based
penalty when delivery falls below that threshold. Most of the remaining components also focus
on penalizing readily observable undesirable behavior, including persistent delivery
failures, invalid messages, excessive IP colocation, and protocol violations. The
application-specific score provides an additional deployment-dependent contribution whose
meaning is defined outside the general GossipSub protocol.

The counters associated with scoring terms decay periodically by multiplication with
configurable decay factors. When a decaying counter falls below the configured zero
threshold, the protocol treats it as zero. Decay makes recent behavior more influential than
sufficiently old behavior and prevents a finite contribution from remaining in the score
indefinitely.

\subsubsection{Opportunistic Grafting}

Opportunistic grafting helps a node recover when its mesh peers perform poorly even though
their number remains within the configured bounds. At configured intervals, the node compares
the median score of its mesh peers with the opportunistic-grafting threshold. If the median
is below this threshold, the node grafts additional eligible peers whose scores exceed the
current mesh median. These new peers give the node an opportunity to improve the quality of
its mesh peers without first waiting for existing peers to obtain negative scores or for their
number to fall below $D_\text{lo}$.

\subsubsection{Heartbeat and Meta Peers}

GossipSub performs mesh maintenance periodically through a heartbeat executed independently
by each node. During a heartbeat, the node removes ineligible mesh peers, adjusts the number
of mesh peers toward $D$, enforces the outbound quota, and performs opportunistic grafting
when its own interval expires.

The heartbeat also disseminates metadata about recently seen messages to selected eligible
peers that are not mesh peers of the node. These peers are selected from the locally known set
of topic participants, and we refer to this set as the node's meta peers. Rather than sending
each full message eagerly to such a peer, a node advertises available message identifiers in
an \texttt{IHAVE} control message. A peer that receives an \texttt{IHAVE} message may request
identifiers it has not seen by returning an \texttt{IWANT} control message, after which the
advertising peer sends the requested full messages if they remain available. This metadata
layer provides an additional dissemination path outside the mesh and is therefore distinct
from the mesh-based forwarding behavior on which this chapter focuses.

\subsection{Goal}

Our objective is to violate the liveness of message delivery through one GossipSub mesh by
separating a constant fraction of participating nodes from honest mesh propagation. We say
that a target node is separated from the mesh when every one of its mesh-peer positions is
occupied by an attacker-controlled peer. After achieving this state, the adversary controls
which mesh messages are delivered to the target and may withhold messages that originate from
honest nodes. This notion of separation concerns delivery through mesh peers and does not
exclude delivery through the metadata exchange described above.

\subsection{Challenges}

To occupy all mesh-peer positions of a target, attacker peers must obtain sufficiently high
peer scores to be retained when the target prunes its original honest peers based on peer
scoring. Among the scoring components, First Message Deliveries is both the most important
positive contribution and a directly manageable term through message-delivery timing. The
principal challenge is therefore to make attacker peers deliver a sufficiently large fraction
of messages before the competing honest peers.

As in our attack on DOG (Section~\ref{sec:dog}), an adversary able to publish fresh valid
messages at an arbitrarily high frequency could send those messages through its own peers
first and thereby obtain high first-delivery scores relatively easily. However, we may not be
able to rely on this capability because the higher-level application may restrict the
production rate of valid messages. For example, frequently producing valid Ethereum consensus
votes requires control of many eligible validators and their corresponding voting power.

We also do not assume control over honest nodes, honest links, or the placement and
connectivity of the honest network, because such control would give the adversary an
artificial delivery advantage inconsistent with the protocol model. The main challenge is
therefore to obtain a systematic first-delivery advantage for messages originating throughout
an otherwise uncontrolled honest network while using only the attacker-controlled
infrastructure introduced later.

\subsection{Basic Attack Model}
\label{sec:gossipsub-model}

We model the honest network participants as $n$ nodes $X = \{x_1, \dots, x_n\}$ in a planar
space $[0,1]^2$ with a large $n$. We model the latency between two nodes $u$ and $v$ as the
$L_2$ Euclidean distance $d(u,v) = \lVert u - v \rVert_2$ between their positions in the
plane. We assume the honest nodes are i.i.d.\ sampled from a uniform distribution
$\mathcal{D}$ over the space $[0,1]^2$. We model the messages as being sourced uniformly from
all the $n$ honest nodes.

We assume that the attacker can deploy $k$ attacker nodes $r_1, \dots, r_k$ in the same
space, with $k$ being some affordable constant. We assume that the attacker has better network
service than the honest links, represented by a latency factor of $q \in [0,1]$. Formally, the
internal latency between two attackers $r_i$ and $r_j$ is $d'(r_i, r_j) = q \cdot d(r_i,
r_j)$. In a simplified analysis, we may assume the attackers are also i.i.d.\ sampled from the
same uniform distribution $\mathcal{D}$.

As an extension, we can also consider additional highly clustered nodes in a constant number
of clusters.

As an additional assumption only for attacking the outbound-connections quota, we assume
that the candidate pool of a node, that is, its meta peers, is sampled uniformly from all
network nodes. We also assume that the attacker controls a small constant fraction (for
example, on the level of 2\%) of the identities in the pool.

\subsubsection{Explaining the Model}

We model the nodes on a plane with Euclidean distances in order to model the geographical
latency between nodes in reality. We use the $L_2$ distance to simplify the analysis compared
to using real-world geographical distances or considering Riemannian geometry on a ball,
which would be more complicated. As this is a theoretical analysis, we expect this
simplification not to affect the general magnitude of the results.

We assume a better network service as a capability of the attacker. This may be achieved by
using some satellite network service, such as DoubleZero~\cite{doublezero}, to achieve a
better latency than the typical geographical links. Due to the value of the attack, the
attacker could be more incentivized to pay for a better network service than the typical
honest links.

We model the distribution of nodes as a uniform distribution with additional clustered
nodes, in order to model standalone participants and large data centers, respectively. Since
the number of large data centers is not too high, we model it as an affordable constant for
the attacker.

\subsection{Attacking the First Message Deliveries Peer Scoring}

The most important part of the attack is to control the First Message Deliveries term of the
peer scoring. Therefore, we first analyze the methods to take over the entire score-based
mesh-peer quota ($D_\text{score}$) of a target node $u$.

\subsubsection{Condition Analysis}

We say that an attacker node $r_i$ can control a message from $v$ to $u$ via the mesh peer
$p$ of $u$ if the attacker can deliver the message to $u$ via $r_i$ before peer $p$ does:
$$
  I^{(i)}_{u,p,v} := \mathbb{1}\left[\min_j \Bigl( d(v, r_j) + q \cdot d(r_j, r_i) + d(r_i, u)
  \Bigr) < d(v, p) + d(p, u)\right]\text{.}
$$

Since there are $n$ possible peers $p$, a sufficient condition for controlling a message
from $v$ to $u$ via any peer $p$ is:
$$
  I^{(i)}_{u,v} := \mathbb{1}\left[\min_j \Bigl( d(v, r_j) + q \cdot d(r_j, r_i) + d(r_i, u)
  \Bigr) \le d(v, u)\right]\text{.}
$$

Let the control rate of a node $u$ be the expected rate of messages from all sources to $u$
that are controlled, via any peer $p$, by an optimal $r_i$, under the conditions above:
$$
  \alpha_u := \max_i \frac{1}{n}\sum_{v \in X} I^{(i)}_{u,v}
  \approx \max_i \mathbb{E}_{v\sim\mathcal{D}}\left[I^{(i)}_{u,v}\right]\text{.}
$$

Similarly, let the control rate of a possible mesh link $u \gets p$ be
$$
  \alpha_{u,p} := \max_i \frac{1}{n}\sum_{v \in X} I^{(i)}_{u,p,v}
  \approx \max_i \mathbb{E}_{v\sim\mathcal{D}}\left[I^{(i)}_{u,p,v}\right]\text{.}
$$

A sufficient condition for separating a node $u$ from the network among the peers selected
by the First Message Deliveries peer scoring is $\alpha_u > T$, with the threshold $T =
\frac{D_\text{score}}{D_\text{score}+1}$. In a practical attack, we may need a rate slightly
higher than this, for example with $T = \frac{(1+\epsilon)D_\text{score}}{(1+\epsilon)
D_\text{score}+1}$.

To achieve the separation, the attacker can have $r_i$ establish $D_\text{score}$ links to
$u$ and can then control messages from all sources to $u$ via these $D_\text{score}$
connections (with different IPs, assuming this is cheap). Each of these links is expected to
have a peer score fraction of $> \frac{1}{D_\text{score}+1}$ or $>
\frac{1+\epsilon}{(1+\epsilon)D_\text{score}+1}$, while all the other peers together only have
a fraction of $< \frac{1}{D_\text{score}+1}$ or $< \frac{1}{(1+\epsilon)D_\text{score}+1}$,
which is not enough to be selected by the First Message Deliveries peer scoring. Therefore,
the attacker links will always be preserved in pruning, while the honest links will eventually
all be pruned by peer scoring.

For fixed $k$, $q$, and attacker locations $r_1, \dots, r_k$, the probability that a
uniformly sampled target satisfies this condition is
$$
  \mathbb{P}\left[\alpha_u > T\right] \approx \mathbb{P}_{u\sim\mathcal{D}}\left[\max_i
  \mathbb{E}_{v\sim\mathcal{D}}\left[I^{(i)}_{u,v}\right] > T\right]\text{.}
$$
The right-hand side does not depend on $n$, so any positive value implies that a constant
fraction of nodes satisfies the sufficient condition for full separation as $n$ grows.

In a similar way, a sufficient condition for being able to remove a single mesh link $p \to
u$ by some $r_i$ is
$$
  \max_i \mathbb{E}_v\left[I^{(i)}_{u,p,v} \mid \text{messages from $v$ are not delivered
  first by other mesh peers of $u$}\right] > T\text{,}
$$
which has a sufficient condition
$$
  \beta_{u,p} > T, \quad \text{with } \beta_{u,p} := \frac{\alpha_u}{\alpha_u + (1 -
  \alpha_{u,p})}\text{.}
$$
The bound $\beta_{u,p}$ is given by the fact that $p$ delivers at most $(1 - \alpha_{u,p})$ of
the messages, while the attacker $r_i$ can deliver at least $\alpha_u$ of the messages. In
this way, the attacker can apply a similar method to prune the honest link $p \to u$ by
establishing $D_\text{score}$ connections with higher peer scores.

For fixed $k$, $q$, and attacker locations, $\mathbb{P}[\beta_{u,p} > T]$ is likewise
independent of $n$ in the spatial model. We expect $\mathbb{P}[\beta_{u,p} > T]$ to be
substantially larger than $\mathbb{P}[\alpha_u > T]$ because separating a target according to
$\alpha_u$ requires the attacker to outperform every possible honest forwarding position,
whereas removing a particular link according to $\beta_{u,p}$ requires it to outperform only
the selected peer $p$. Since a randomly selected peer is generally not positioned optimally
for message forwarding from all message sources, removing a random honest link is easier than
fully separating a target node.

\subsubsection{Numerical Results}

We numerically evaluate the distributions of $\alpha_u$ and $\beta_{u,p}$ under the uniform
spatial model. We sample $n = 1000$ honest nodes and $100$ attacker nodes independently and
uniformly from $[0,1]^2$. For every $k \in \{1, \dots, 100\}$, we use the first $k$ sampled
attacker nodes and calculate $\alpha_u$ for every $u \in X$ and $\beta_{u,p}$ for every
ordered pair $(u,p) \in X^2$. We repeat these computations for latency factors $q \in \{0,
0.5, 0.8\}$, where a smaller $q$ represents a stronger latency advantage for the attacker
network. The codebase is available at~\cite{li-code}.

\begin{figure}[tbp]
  \centering
  \includegraphics[width=\textwidth]{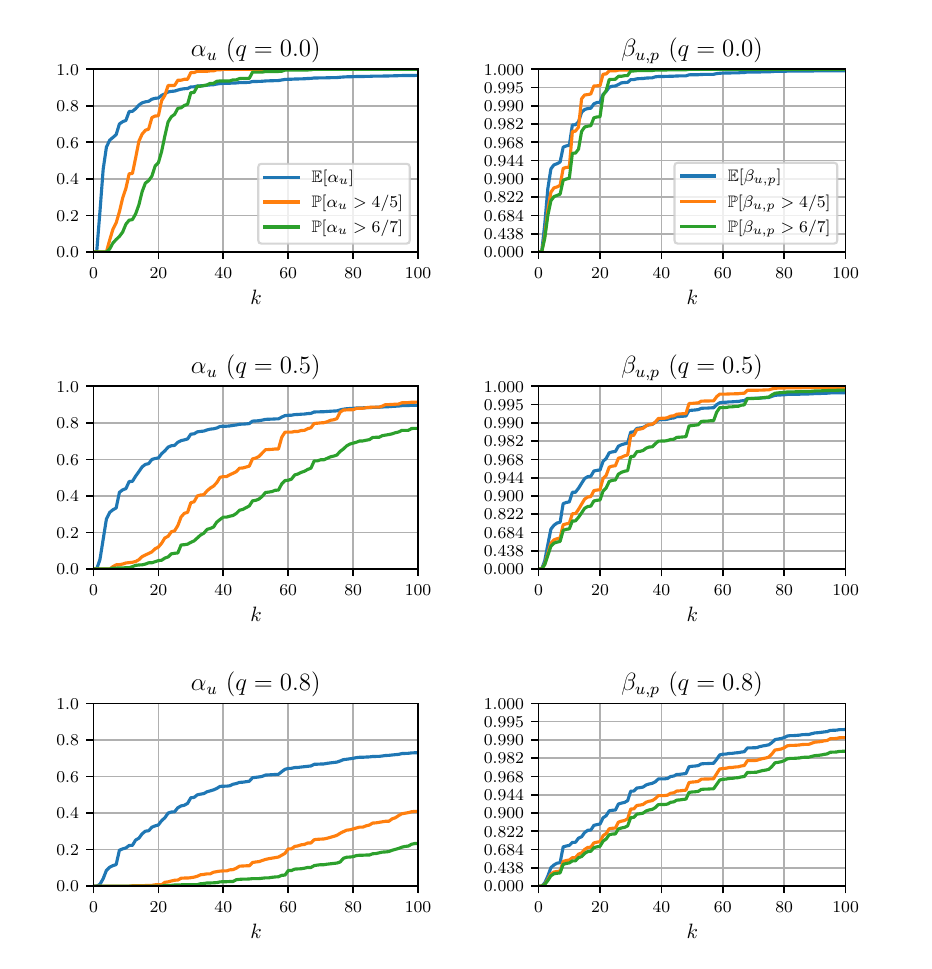}
  \caption{$\mathbb{E}[\alpha_u]$, $\mathbb{E}[\beta_{u,p}]$, $\mathbb{P}[\alpha_u > T]$, and
    $\mathbb{P}[\beta_{u,p} > T]$ for $T \in \{4/5, 6/7\}$ and different values of $q$ and
    $k$.}
  \label{fig:first-delivery-rates}
\end{figure}

Figure~\ref{fig:first-delivery-rates} reports $\mathbb{E}[\alpha_u]$,
$\mathbb{E}[\beta_{u,p}]$, $\mathbb{P}[\alpha_u > T]$, and $\mathbb{P}[\beta_{u,p} > T]$ for
$T = 4/5$ and $T = 6/7$. Under the default value $D_\text{score} = 4$, these thresholds
correspond respectively to the basic condition $\epsilon = 0$ and the more conservative choice
$\epsilon = 0.5$ in the threshold $T = \frac{(1+\epsilon)D_\text{score}}{(1+\epsilon)
D_\text{score}+1}$ defined above.

For fixed attacker locations $r_1, \dots, r_k$, the values calculated over all $u \in X$ and
$(u,p) \in X^2$ approximate the corresponding expectations and probabilities over $u, p \sim
\mathcal{D}$, given these attacker locations. Although we use $n = 1000$ for the numerical
evaluation, the corresponding limiting quantities are independent of $n$, as established in
the condition analysis above. For example, when $q = 0.8$ and $k = 100$, approximately 20\% of
the nodes satisfy $\alpha_u > 6/7$. For the same parameters, approximately 98\%--99\% of the
ordered pairs satisfy $\beta_{u,p} > 6/7$.

\subsubsection{An Iterative Optimization}

Once a node has been separated, all of its mesh peers are controlled by the attacker. We can
therefore assume that messages originating from a separated node are routed immediately
through an attacker peer, without requiring the attacker network to win the geographical
latency race represented by $I^{(i)}_{u,v}$. These additional controlled messages can increase
the control rates of other targets and thereby make further nodes satisfy the separation
condition.

We model this optimization as an iterative process. Let $S_t \subseteq X$ denote the set of
nodes separated after iteration $t$, with $S_0 = \emptyset$, and define the updated control
rate
$$
  \alpha_u(S_t) := \max_i \frac{1}{n}\sum_{v \in X} \mathbb{1}\left[I^{(i)}_{u,v} = 1 \ \lor\
  v \in S_t\right]\text{.}
$$
At each iteration, we update the separated set according to
$$
  S_{t+1} := S_t \cup \{u \in X : \alpha_u(S_t) > T\}\text{.}
$$
The first update $S_1$ is exactly the set obtained by the non-iterative condition $\alpha_u >
T$. Because $S_t \subseteq S_{t+1} \subseteq X$, the process is monotone and reaches a fixed
point after at most $n$ iterations that enlarge the separated set.

\begin{figure}[tbp]
  \centering
  \includegraphics[width=\textwidth]{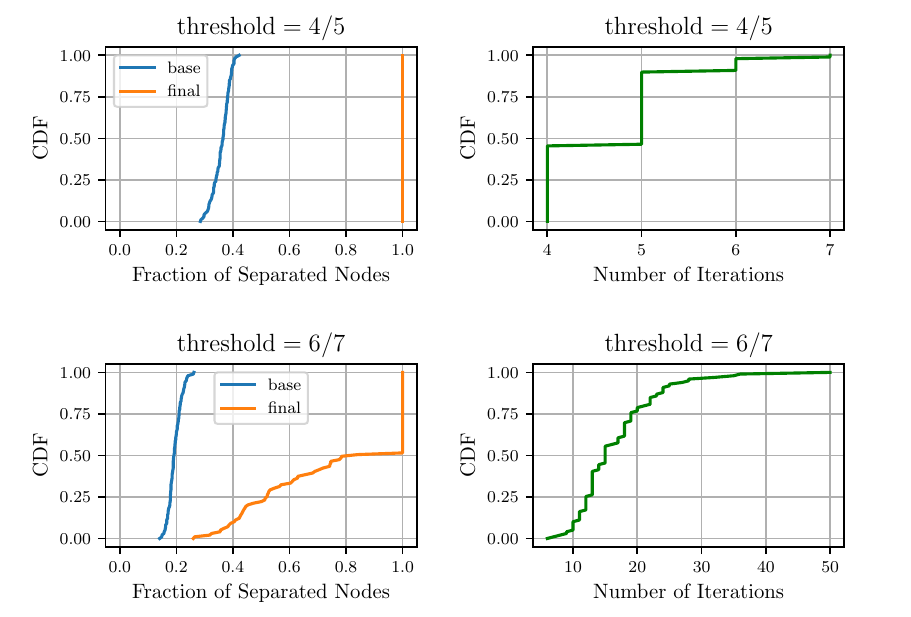}
  \caption{Empirical cumulative distributions of the initial and final fractions of separated
    nodes and of the number of iterations to convergence, over 100 independent runs with $n =
    1000$, $k = 100$, and $q = 0.8$.}
  \label{fig:iterative-separation}
\end{figure}

We evaluate this process over 100 independent runs with $n = 1000$, $k = 100$, and $q =
0.8$, resampling the honest and attacker locations in each run. For each threshold,
Figure~\ref{fig:iterative-separation} compares the empirical cumulative distribution functions
of the initial fraction $|S_1|/n$ and the final fraction at the fixed point, and also reports
the distribution of the number of iterations that enlarge the separated set. For $T = 4/5$,
the initial separated fraction is approximately 40\%, whereas the iteration reaches 100\% in
all 100 runs. For the more conservative threshold $T = 6/7$, iteration still substantially
increases the separated fraction, and approximately half of the runs reach 100\% separation.

We do not rely on this iterative optimization in the main attack analysis, which requires
only that the non-iterative attack separate a positive constant fraction of nodes.
Nevertheless, the simulations indicate that iteration can substantially increase this fraction
and can achieve complete separation for some parameter choices and sampled configurations. The
method therefore appears to offer significant potential for strengthening the attack, although
a general analysis of its final fixed point is beyond the scope of this report.

\subsubsection{Addressing Additional Clustered Nodes}

The attacker can address the additional clustered nodes by deploying one attacker node in
each cluster, since the number of clusters is some affordable constant. Each such attacker has
very low latency to the honest nodes in its cluster and can relay their messages to the other
attackers through the low-latency internal network. The attack can therefore achieve an even
stronger delivery advantage for clustered nodes than for uniformly distributed nodes.

\subsection{Addressing Other Mechanisms}

\subsubsection{Randomly Selected Peers}

Controlling the $D_\text{score}$ score-selected positions does not immediately control the
remaining mesh-peer positions, because GossipSub selects the other peers retained during
pruning randomly. While the mesh remains within its configured bounds, up to $D_\text{hi} -
D_\text{score}$ peers may therefore remain outside the score-protected attacker set.

To replace these peers, the attacker can use additional eligible identities to graft the
target repeatedly. Once the target has more than $D_\text{hi}$ mesh peers, its next pruning
step preserves the high-scoring attacker peers and resamples the remaining retained positions
from the other eligible mesh peers. Each repetition gives attacker identities another
opportunity to occupy positions that were previously assigned to honest peers. Provided that
the attacker has sufficiently many distinct identities, including enough IP addresses to avoid
the IP-colocation penalty, repeating this process can eventually replace all randomly retained
honest peers with attacker-controlled peers.

\subsubsection{Outbound Connections Quota}

The attacks described above establish attacker-controlled inbound connections by actively
grafting the target node and therefore do not directly address the outbound-connections quota.
Consequently, each target may retain up to $D_\text{out}$ honest outbound mesh peers even after
its other mesh-peer positions have been occupied by the attacker.

\paragraph{Consequence of Maintaining Outbound Connections.}
The outbound connections alone may preserve connectivity among the honest nodes, even under
the default value $D_\text{out} = 2$.

To illustrate this consequence, suppose that every node independently selects exactly $d =
D_\text{out}$ outbound peers uniformly from all $n$ nodes, and consider the undirected graph
obtained by ignoring the directions of these connections. If this graph is disconnected, then
there exists a nonempty proper subset $S \subset X$ with no outbound connection crossing
between $S$ and $X \setminus S$ in either direction. Applying a union bound over all nonempty
proper subsets gives
$$
  \mathbb{P}[\text{the outbound graph is disconnected}] \le \sum_{s=1}^{n-1} \binom{n}{s}
  \left(\frac{s}{n}\right)^{ds} \left(\frac{n-s}{n}\right)^{d(n-s)} \xrightarrow{n\to\infty} 0
$$
for any constant $d \ge 2$. Thus, under uniform independent selection, preserving only the
outbound connections is sufficient to keep the induced graph connected with high probability,
so the basic attack does not establish separation of its targets from the mesh. Note that this
conclusion could be different from the one obtained with the $G(n,p)$ Erd\H{o}s--R\'enyi random
graph model, where the threshold for connectivity is $p = \frac{\log n}{n}$.

\paragraph{Attacking Outbound Connections.}
An outbound connection from a target $u$ to an honest peer $p$ is an inbound connection from
the perspective of $p$. The attacker can therefore apply the single-link attack analyzed above
to the connection at $p$ and, if the attack causes the connection to be removed, force $u$ to
sample a replacement outbound peer. This step requires removing only one honest link rather
than fully separating $p$, and the numerical results show that the corresponding condition
$\beta_{u,p} > T$ holds for a much larger fraction of pairs than the full-separation condition
$\alpha_u > T$.

Suppose that attacker identities constitute a fraction $\rho$ of the candidate pool and that
a fraction $h$ of the pool consists of honest nodes whose resulting links the attacker cannot
remove. Candidates corresponding to removable honest links lead to another replacement
attempt, so the eventual outcome is determined by whether the target first selects an attacker
identity or an honest node whose link cannot be removed. The probability that one outbound
position is eventually occupied by the attacker is therefore approximately
$$
  p_\text{out} \approx \frac{\rho}{\rho + h}\text{.}
$$
For example, if attacker identities constitute approximately 2\% of the pool and only
approximately 2\% of the pool produces honest links that cannot be removed, then
$p_\text{out} \approx 0.5$. Because the implementation samples each replacement peer
independently, the probability of occupying all $D_\text{out}$ positions is approximately
$p_\text{out}^{D_\text{out}}$, which is $0.25$ for the default $D_\text{out} = 2$ in this
example. This success probability is a positive constant independent of $n$.

The calculation relies on the additional assumptions in Section~\ref{sec:gossipsub-model}
that the candidate pool is sampled uniformly from all network nodes and that attacker
identities constitute a small constant fraction $\rho$ of the pool, for example $\rho = 2\%$.
Uniform sampling is needed to apply the single-link result above to replacement candidates:
when $k = 100$, $q = 0.8$, and $T = 6/7$, only approximately 1\%--2\% of uniformly sampled
pairs fail to satisfy $\beta_{u,p} > T$, giving $h \approx 1\%\text{--}2\%$. It also requires
that an attacker peer, once selected, remains eligible and is retained by behaving according
to the protocol.

\subsubsection{Time in Mesh Peer Scoring}

We do not focus on the Time in Mesh component as we expect First Message Deliveries to be
more important and more directly controllable. Nevertheless, the attacker can also address
this component in several ways.

The attacker can introduce a small initial set of peers and allow them to remain in the
target's mesh long enough to accumulate Time in Mesh credit before triggering pruning with
additional attacker connections. It can then enlarge this set gradually, allowing each new
group of attacker peers to accumulate credit before the next stage. High First Message
Deliveries scores can further compensate for lower Time in Mesh scores, and initially using
fewer attacker peers can further concentrate controlled first deliveries to compensate the
nodes with lower scores.

\subsection{Details of Performing and Maintaining the Attack}

\subsubsection{Time Cost}

Fully separating a target node requires occupying both its $D_\text{score}$ score-selected
mesh-peer positions and its randomly retained positions, and these two parts of the attack
proceed through different mechanisms with different time costs.

Occupying the score-selected positions requires the attacker's peers to accumulate a
sufficient First Message Deliveries advantage, since this term decays and accumulates
gradually over time. Under the default decay parameters, reaching a peer score high enough to
secure an $\epsilon = 50\%$ advantage over competing honest peers takes roughly less than 3
minutes, even if a competing honest peer already has a high initial score. Occupying the
randomly retained positions instead requires repeatedly grafting the target to trigger
pruning, as described above, which can be performed much faster given enough attacker
identities to use; this can be done in roughly less than 30 seconds. Therefore, the total time
to perform the initial attack on one target node is roughly less than 3.5 minutes. The attack
can be performed independently on multiple target nodes in parallel, so the total time to
separate a constant fraction of nodes is also roughly less than 3.5 minutes.

To maintain separation, the attacker must apply the same methods whenever a new honest peer
is grafted, for example after churn. Since a newly grafted honest peer starts without an
established peer score, it does not require waiting through the long decay period needed to
overcome an established competing peer, so replacing it takes roughly less than 30 seconds.

\subsubsection{Minimal Message Delivery}

The Mesh Message Deliveries term of peer scoring penalizes a mesh peer whose delivery rate
stays below a configured threshold after its activation period. In addition, opportunistic
grafting will graft new peers if the median peer score of the mesh drops below a threshold.
Therefore, once the attacker has taken over the links of a target node, it must still maintain
a minimal level of message delivery to avoid being penalized or replaced by this mechanism.

The Mesh Message Deliveries component contributes no penalty as long as the decaying delivery
counter of a peer stays at or above the configured threshold, so it is sufficient for the
attacker to relay a selected subset of the messages at the required rate and to withhold the
remaining ones. Since the counter also credits deliveries that arrive within a configured
window after the first delivery, the attacker can forward each selected message over all of
its connections to the target within that window. Every attacker connection therefore
satisfies its own threshold from the same relayed subset, and the delivery cost does not scale
with the number of mesh-peer positions occupied at the target.

Maintaining this minimal delivery also keeps the peer scores of the attacker connections
above some basic level, since the same relayed messages continue to contribute to their First
Message Deliveries counters against the ongoing decay. The median score of the target's mesh
can therefore stay above the opportunistic-grafting threshold, so the target does not graft
additional peers on that basis.

\subsubsection{Considering Churn}

Since the attack is performed independently on each target node, a node joining the topic is
a new target that the attacker can separate independently, following the same procedure as in
the initial attack. The attacker can likewise repair the separation of an individual node
independently whenever that node is disturbed by churn. A node joining the topic grafts $D$
peers, so a single join interferes with at most $D$ already separated nodes, and the
interference on each of them consists of one newly grafted honest peer that must be replaced.
As established above, replacing such a peer takes at most $\tau \approx 30$ seconds.

Let $c$ denote the fraction of the network that joins the topic within one interval of length
$\tau$. Each such join disturbs at most $D$ separated nodes, so the expected fraction of
separated nodes disturbed per interval is at most $Dc$, and each disturbance costs at most one
interval of repair. The expected fraction of time for which an already separated node is not
fully separated is therefore at most $Dc$. Requiring this fraction to be at most $1/2$ gives
the condition
$$
  c \le \frac{1}{2D}\text{,}
$$
which under the default value $D = 8$ means a churn of at most 6.25\% of the network joining
per 30 seconds.

This bound is well above the churn rates expected in a deployed network of this kind, so the
attacker can maintain the separation of a constant fraction of nodes for a constant fraction
of the time after the initial attack. Note also that a joining node disturbs at most $D$
separated nodes regardless of $n$, so the condition constrains only the relative churn rate
and remains independent of the network size.

\subsection{Result Summary and Limitations}

Under the geographical model analyzed in this chapter, a constant number of geographically
distributed attacker nodes with the assumed latency advantage can separate a constant fraction
of nodes from the GossipSub mesh, before accounting for the outbound-connections quota. The
attack requires a larger but still constant number of network identities. The separation can
be maintained with limited message delivery, provided that network churn stays below a bound
that is independent of the network size and is well above realistic churn rates.

The outbound-connections quota is not defeated by the basic attack because a target chooses
its own outbound peers. Our extension for attacking these peers assumes that the candidate
pool is sampled uniformly from all network nodes and that attacker identities constitute a
small but sufficient constant fraction of this pool (e.g., 2\% under some reasonable
configurations). Under this assumption, repeatedly causing an honest outbound peer to be
replaced gives an attacker identity a corresponding probability of being selected as its
replacement.

The attack separates only a constant fraction of nodes, while the remaining constant fraction
may still be connected; we do not analyze connectivity among those remaining nodes. The attack
also cannot displace statically configured peers whose retention is not governed by the
peer-selection mechanisms considered here. Finally, delivery from non-mesh peers through
\texttt{IHAVE} and \texttt{IWANT} messages remains possible because the metadata layer and the
formation of its peer set are outside the scope of our analysis.

%% file: sections/dog.tex

\section{Attacking DOG}
\label{sec:dog}

\subsection{The DOG Protocol}

DOG (Dynamic Optimal Graph)~\cite{informal-dog, cometbft-dog-spec} is a transaction
dissemination protocol designed for CometBFT~\cite{cometbft}, the consensus engine used by the
Cosmos ecosystem. Each node maintains a set of persistent connections to peers, and a
transaction published by any node is disseminated by repeated forwarding over these
connections.

DOG reduces the redundant traffic produced by flooding without removing any peer connection.
Instead of selecting which peers to forward to, each node decides, for every pair consisting
of an incoming and an outgoing connection, whether transactions arriving on the former should
be forwarded on the latter. Such a pair is called a route, and a node disables a route once it
learns that forwarding along it produced a redundant delivery. DOG additionally applies a
redundancy control mechanism, which measures how much duplication a node observes and admits
further disabling only while this quantity is above a target range, restoring previously
disabled routes while it is below. The two mechanisms are described in the following
subsections.

\subsubsection{Route Filtering}

The central mechanism of DOG is route filtering, by which a node stops forwarding
transactions along paths that have been observed to deliver redundantly. For an intermediate
node $u$, a route is an ordered pair $(p, t)$ of neighbors of $u$, representing the propagation
path $\cdots \to p \to u \to t$ in which $u$ receives a transaction from $p$ and forwards it to
$t$. Node $u$ forwards a transaction received from $p$ to $t$ only if the route $(p, t)$ is not
filtered by $u$. Routes are directional and are maintained independently of the underlying
connections, so filtering $(p, t)$ neither removes the connection between $u$ and $t$ nor
affects any other route of $u$ involving $p$ or $t$.

A node filters a route when it is informed that one of its own forwardings was redundant.
When a node $t$ receives a transaction that it has already received from elsewhere, it may
return a \texttt{haveTx} message to the sender $u$ identifying that transaction. Upon receiving
this message, $u$ looks up the peer $p$ from which it originally received the transaction and
filters the route $(p, t)$, so that transactions subsequently arriving from $p$ are no longer
forwarded to $t$. Route filtering itself provides no means of restoring a filtered route, so
the route remains filtered unless the involved connection is re-established or another
mechanism restores it.

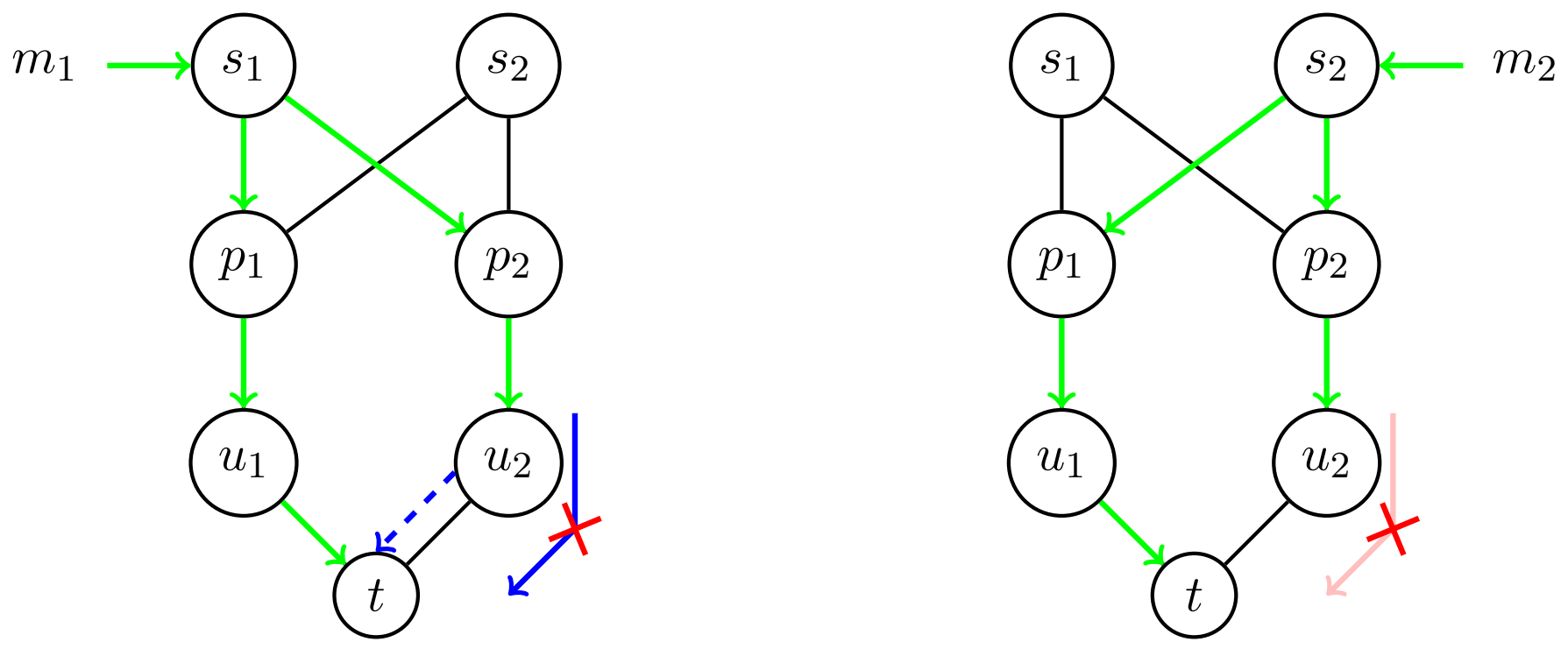
\begin{figure}[htb]
  \centering
  \begin{subfigure}{0.45\textwidth}
    \centering
    \input{figures/graphs/route-filtering-1.tex}
  \end{subfigure}
  \begin{subfigure}{0.45\textwidth}
    \centering
    \input{figures/graphs/route-filtering-2.tex}
  \end{subfigure}
  \caption{An example of route filtering.}
  \label{fig:route-filtering}
\end{figure}

Figure~\ref{fig:route-filtering} shows an example of route filtering. First, $s_1$ publishes a
transaction $m_1$, which reaches $t$ via $u_1$. The same transaction later also reaches $t$ via
$u_2$, and since this delivery is redundant for $t$, the route $p_2 \to u_2 \to t$ is filtered
by $u_2$. When $s_2$ subsequently publishes a transaction $m_2$, the filtered route is no
longer taken, and $m_2$ reaches $t$ via $u_1$ only.

\subsubsection{Redundancy Control}

Redundancy control regulates route filtering in both directions, so that the amount of
duplicate delivery observed by a node stays within a configured target range. Two parameters
govern the mechanism: a redundancy control interval, set to 1 second by default, and a target
redundancy range, set to $1.0 \pm 10\%$ by default. At the end of each interval, a node
computes its redundancy level over that interval as
$$
  \text{redundancy level} = \frac{\text{number of redundant transactions received in the
  interval}}{\text{number of new transactions received in the interval}}\text{,}
$$
and compares it with the target range.

If the level exceeds the upper bound, the node grants itself permission to send one
\texttt{haveTx} message; this permission is consumed by the first redundant transaction it
receives afterwards, and the resulting filtering removes one route and thereby lowers its
redundancy. If the level falls below the lower bound, the node instead sends a
\texttt{resetRoute} message to a randomly selected neighbor, which re-enables one randomly
chosen filtered route of that neighbor and thereby raises its redundancy. If the level lies
within the target range, or if the node received no transactions during the interval, it takes
no action. Since at most one such permission is outstanding at any time, a node causes at most
one route to be filtered per interval. The actions taken at the end of an interval are
summarized as follows:
$$
  \begin{cases}
    \text{if no transactions are received}          & \implies \text{do nothing}                               \\
    \text{if redundancy level} > \text{upper bound} & \implies \text{allow sending next \texttt{haveTx} (once)} \\
    \text{if redundancy level} < \text{lower bound} & \implies \text{send \texttt{resetRoute}}                  \\
    \text{otherwise}                                & \implies \text{do nothing}
  \end{cases}
$$

\subsection{Overview}

For DOG to retain the delivery guarantee that flooding provides, the effective dissemination
graph formed by the unfiltered routes has to remain connected. We did not find an intuitive
argument establishing that route filtering preserves this property, and we instead obtained
counterexamples showing that DOG may fail to preserve connectivity even in the absence of
failures or Byzantine participants. We then turn to an active adversary and present a practical
attack that separates a full network. Finally, we analyze the cost of the attack and discuss
the assumptions and conditions under which it works, together with potential extensions and
fixes.

The counterexamples show that route filtering alone can disconnect the dissemination graph. A
ring of five nodes is already sufficient when the redundancy is zero, and the construction
generalizes to any target redundancy level. In both cases, redundancy control does not restore
the lost connectivity, because the nodes that would have to request a reset no longer receive
the transactions that would trigger it.

The practical attack requires only a single physical attacker node holding a constant number
of network identities, and it requires no control over honest nodes or over the network
topology. Its essential ingredient is that a participant of a P2P network may publish its own
transactions directly to any chosen nodes and may delay them arbitrarily, which is sufficient
to make an attacker-controlled node the fastest forwarder to a target and thereby have the
competing routes filtered. Applying this procedure to every route disables all forwarding in
the network: a transaction still reaches the direct neighbors of its publisher, but no node
forwards it any further. The attacker can maintain this state indefinitely at a cost
proportional to the traffic published in the network, and the time needed to attack a single
route matches the lower bound imposed by the redundancy control interval.

\subsection{Theoretical Counterexamples}

\subsubsection{A Simple Ring Example}

DOG was expected to preserve transaction delivery with zero redundancy in non-Byzantine
networks, as the resulting dissemination graph was expected to be a superposition of directed
spanning trees. However, the following simple counterexample shows that the filtered
dissemination graph can become disconnected even in the absence of failures or Byzantine
participants.

\begin{figure}[htb]
  \centering
  \begin{subfigure}{.3\textwidth}
    \centering
    \input{figures/graphs/simple-ring-example-1.tex}
  \end{subfigure}
  \hfill
  \begin{subfigure}{.3\textwidth}
    \centering
    \input{figures/graphs/simple-ring-example-2.tex}
  \end{subfigure}
  \hfill
  \begin{subfigure}{.3\textwidth}
    \centering
    \input{figures/graphs/simple-ring-example-3.tex}
  \end{subfigure}
  \caption{The counterexample of a simple ring.}
  \label{fig:simple-ring-example}
\end{figure}
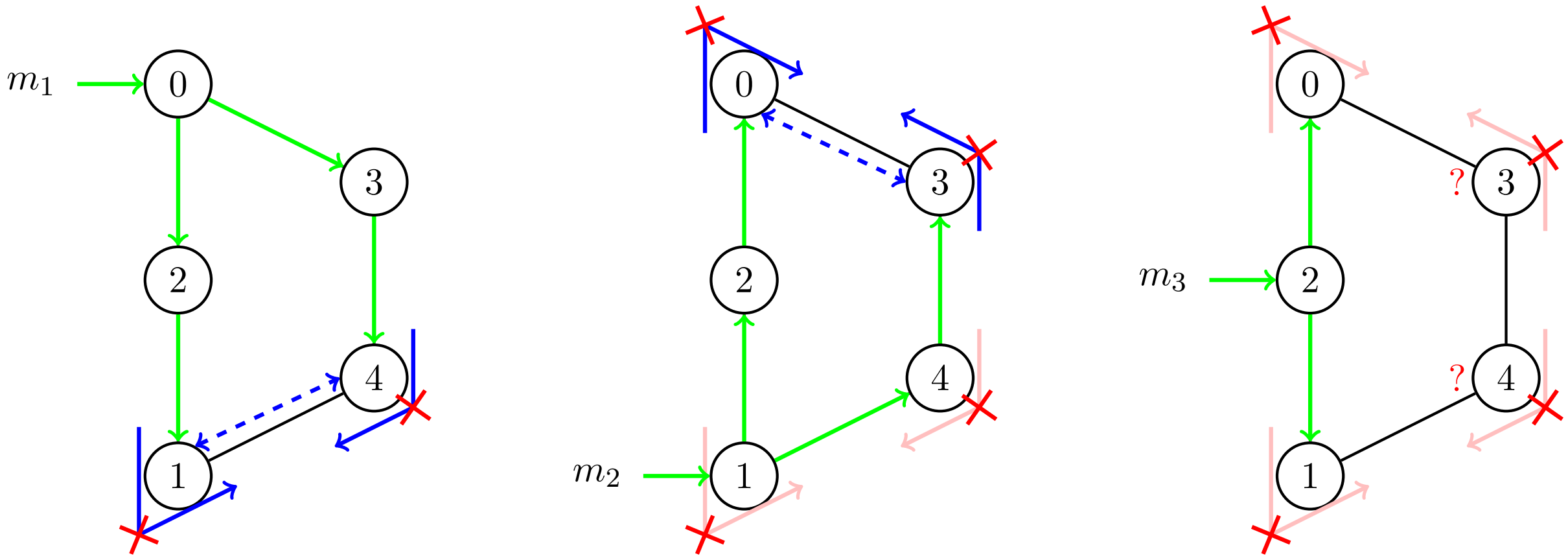

The example is shown in Figure~\ref{fig:simple-ring-example}. First, node $0$ publishes $m_1$,
which reaches nodes $1$ and $4$ simultaneously. As a result, routes $3 \to 4 \to 1$ and $2 \to
1 \to 4$ are filtered because they deliver $m_1$ redundantly. Next, node $1$ similarly
publishes $m_2$, causing routes $4 \to 3 \to 0$ and $2 \to 0 \to 3$ to be filtered. The
resulting dissemination graph is disconnected: when node $2$ publishes $m_3$, neither node $0$
nor node $1$ forwards it, so nodes $3$ and $4$ do not receive it. Since redundancy is zero in
this scenario, this separation cannot be recovered by the redundancy control mechanism.

\subsubsection{Generalized Example with Redundancy}

The ring example can be generalized to allow a positive target redundancy level. The
construction below uses a target redundancy level of 1.0 and still produces a disconnected
dissemination graph without failures or Byzantine participants. The same construction can be
extended to any target redundancy level.

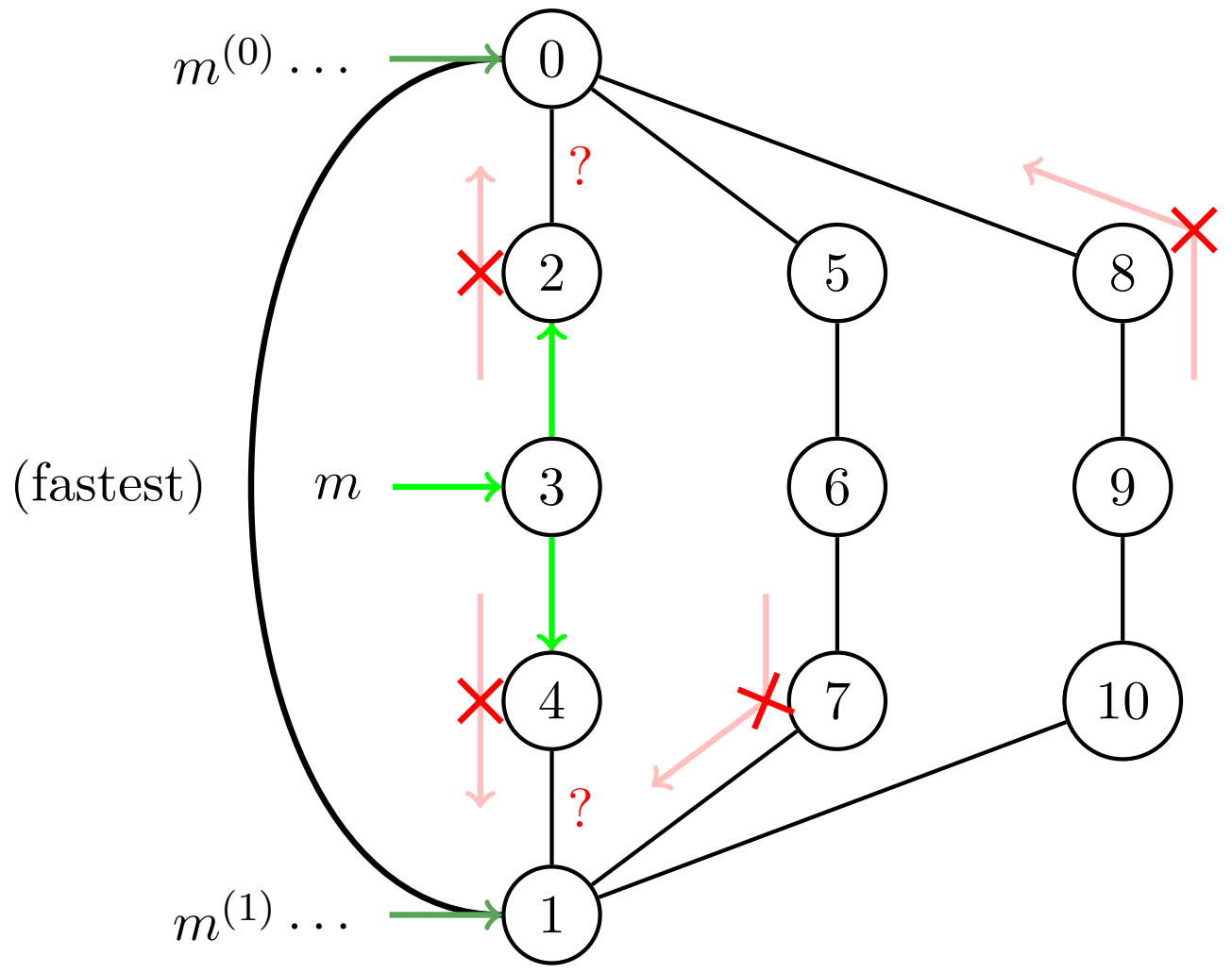
\begin{figure}[htb]
  \centering
  \input{figures/graphs/generalized-redundancy-example.tex}
  \caption{A generalized example with redundancy.}
  \label{fig:generalized-redundancy-example}
\end{figure}

The construction is shown in Figure~\ref{fig:generalized-redundancy-example}. Nodes $0$ and
$1$ initially publish transactions until their redundancy levels stabilize at 1.0. Each node
then retains two unfiltered incoming paths, including the direct path, and causes two of the
other three paths to be filtered at its end. Consequently, at least one of the three indirect
paths is filtered at both ends; without loss of generality, let this be the path $2 - 3 - 4$.
When node $3$ subsequently publishes a transaction $m$, neither node $2$ nor node $4$ forwards
it, so the remaining nodes do not receive $m$. If nodes $0$ and $1$ receive no further
transactions, neither triggers a \texttt{resetRoute}, and the separation persists. For a higher
target redundancy level, adding further paths between nodes $0$ and $1$ yields the same
argument: at least one path is filtered at both ends.

\subsubsection{A Potential Attacking Idea}
\label{sec:potential-attack}

The same filtering behavior also permits an active attack in a restricted scenario. If the
attacker controls every forwarder on the routes that remain unfiltered between a source and a
target, disconnecting those forwarders separates the target from the source.

\begin{figure}[htb]
  \centering
  \input{figures/graphs/potential-attack-example.tex}
  \caption{A potential attack.}
  \label{fig:potential-attack-example}
\end{figure}
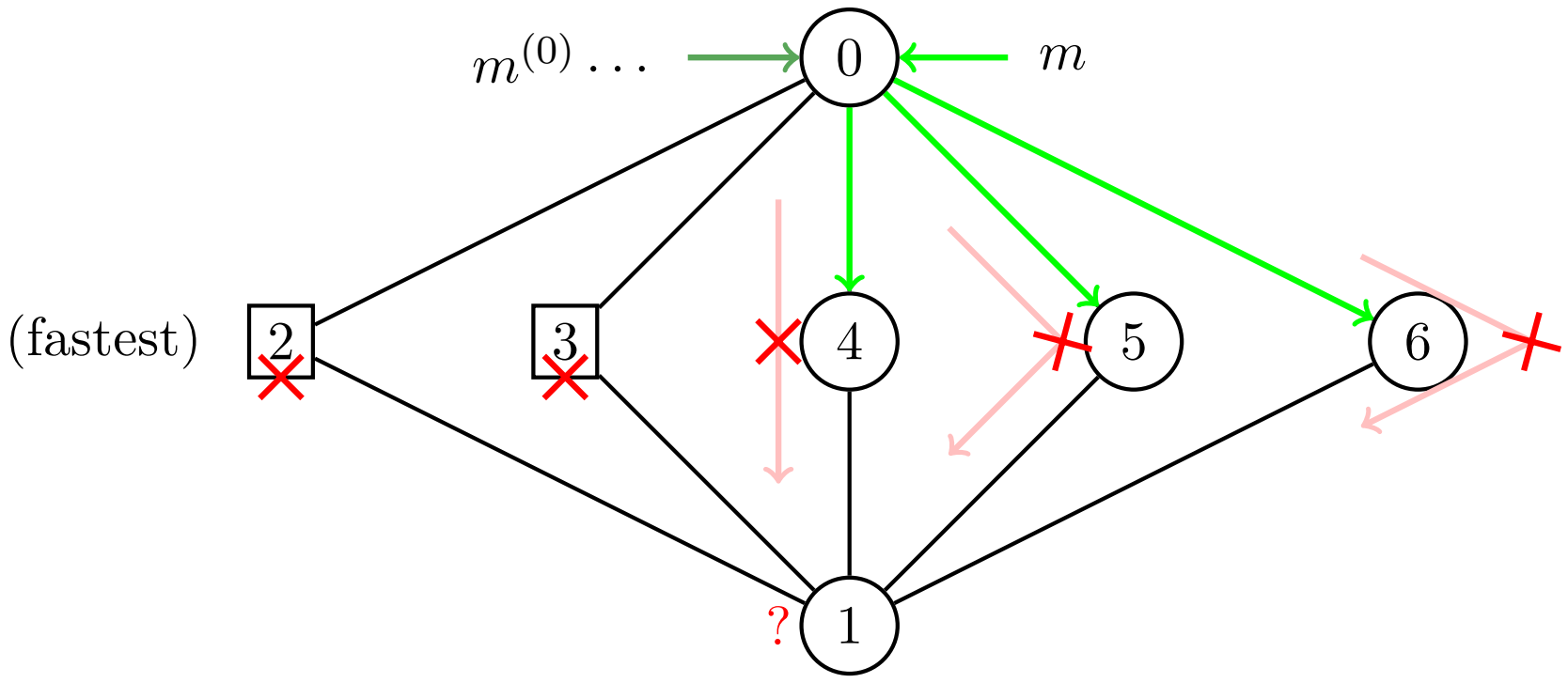

Figure~\ref{fig:potential-attack-example} shows such a scenario. In the scenario, node $0$
keeps publishing transactions. At a target redundancy level of 1.0, node $1$ preserves the
fastest route from node $0$ and one additional route, while causing all other routes from node
$0$ to be filtered. If attacker-controlled nodes $2$ and $3$ are the forwarders on these two
preserved routes, disconnecting them leaves no route by which transactions from node $0$ can
reach node $1$. This separation persists while node $1$ receives no further transactions.

This scenario assumes control over both the network topology and the transaction publication
pattern. The next section generalizes this idea into a practical attack that does not require
such control.

\subsection{A Practical Attack}

\subsubsection{Attacking a Single Route}

\begin{figure}[htb]
  \centering
  \input{figures/graphs/attacking-single-route.tex}
  \caption{Attacking a single route.}
  \label{fig:attacking-single-route}
\end{figure}
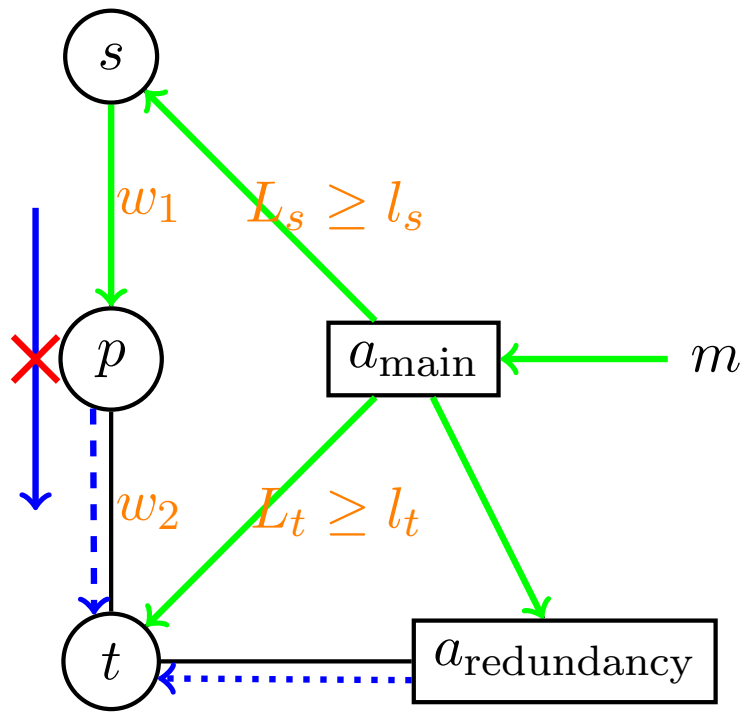

The central requirement of the scenario in Section~\ref{sec:potential-attack} is to make an
attacker-controlled node the fastest forwarder to the target, allowing the competing routes to
be filtered. This can be achieved without controlling the network topology because the attacker
can publish its own transactions and control their initial propagation. This control provides
the propagation-delay advantage needed to make an attacker-controlled node the fastest
forwarder to the target.

The method is shown in Figure~\ref{fig:attacking-single-route}. Suppose the attacker controls
a node $a_\text{main}$ and seeks to filter an active route $s \to p \to t$. Node $a_\text{main}$
publishes a transaction $m$ directly to both $s$ and $t$. The attacker can increase the
propagation delays from $a_\text{main}$ to $s$ and $t$, denoted by $L_s$ and $L_t$, above their
respective lower bounds $l_s$ and $l_t$ imposed by network latency. Choosing these delays such
that $L_t - w_2 < L_s + w_1 < L_t + w_2$ makes $a_\text{main}$ the first node to deliver $m$ to
$t$ and makes $s \to p \to t$ the next route to deliver it, causing that route to be filtered.

To keep the measured redundancy above the target, the attacker uses one or more additional
nodes, denoted by $a_\text{redundancy}$, to forward $m$ from $a_\text{main}$ to $t$. These
deliveries must occur after the delivery through $s \to p \to t$ but within the same redundancy
control interval.

\subsubsection{Separating a Target Node}

The per-route attack can be repeated to filter every forwarding route to a target node and
thereby separate it from the network. Once the target is separated, no other node forwards
transactions to it.

The attacker filters one active route at a time, beginning with the fastest route and
gradually adjusting $L_s$ and $L_t$ to target each successive active route. Throughout this
process, the attacker keeps the redundancy level of $t$ above the target, preventing route
resets and preserving the results of earlier rounds.

One minor edge case occurs when $s - p - t$ forms a triangle, so that the direct connection
$s - t$ exists alongside the route $s \to p \to t$. This case may require additional handling
but does not alter the general method; Section~\ref{sec:dealing-with-triangles} describes its
treatment.

\subsubsection{Attacking a Full Network}

The attacker can separate the full network by applying the target-node attack to every node.
Care is required to avoid restoring routes to a previously separated target when that target
receives a transaction during a later round. Section~\ref{sec:implementation} describes these
cases and how the implementation avoids them.

The attack may not prevent a transaction from being delivered directly from its publisher $s$
to a neighbor $t$, because the attacker cannot control the latency of the link $s - t$.
However, after every target has been separated, no receiving node forwards the transaction
further.

\subsubsection{Maintaining Separation}
\label{sec:maintaining-separation}

For redundancy control to restore a filtered route to a separated node $t$, node $t$ must
first receive a transaction. Since no node forwards transactions to $t$ after the separation,
this can occur only through the direct publisher-to-neighbor delivery described above. Such a
direct delivery may cause $t$ to request a route reset.

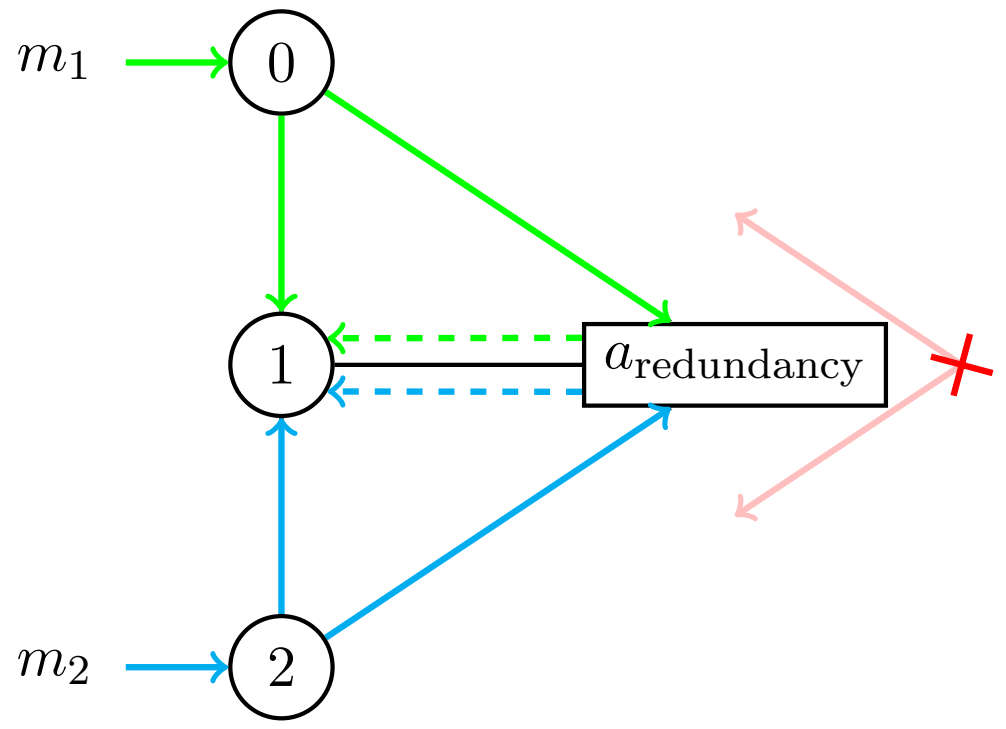
\begin{figure}[htb]
  \centering
  \input{figures/graphs/maintaining-separation.tex}
  \caption{Maintaining separation.}
  \label{fig:maintaining-separation}
\end{figure}

To prevent these resets, the redundancy nodes $a_\text{redundancy}$ listen for transactions
published by each node $s$ and forward each transaction to the direct neighbors of $s$.
Whenever a neighbor $t$ receives a transaction directly from $s$, it therefore also receives
enough redundant deliveries from $a_\text{redundancy}$ to keep its redundancy level above the
target. Figure~\ref{fig:maintaining-separation} illustrates this method.

\subsection{Implementation Details}
\label{sec:implementation}

Implementing the attack requires addressing several details and edge cases. These
considerations do not affect the general method described above and are relevant only to
readers interested in the implementation details.

\subsubsection{Avoiding Interference}

To repeat the per-route or per-node attack in the full network, we need to be careful to
avoid interference. In particular, the filtering state may change between rounds, and a later
round may inadvertently restore a previously filtered route.

Each round assumes that transactions propagate through $a_\text{main} \to s \to u$, but this
route may have been filtered in an earlier round. To restore it, $a_\text{main}$ reconnects to
$s$ before each round, causing the routes $a_\text{main} \to s \to *$ to become active again.
This follows the DOG implementation, which resets filtered routes involving a connection when
that connection is re-established.

We must also avoid inadvertently restoring routes filtered in earlier rounds, which can occur
only if their targets receive a transaction. If a previously filtered route leads to the
current target $t$, this is not a concern because the attack keeps the redundancy level of $t$
above the target range throughout the round. Every other previously filtered route leads to a
target that has already been separated. Since attacker-controlled nodes are not themselves
targets of the separation, such a target may still receive an attacker transaction either
directly from an attacker-controlled node or after one forwarding hop along a route originating
from that node. During the current round, $a_\text{main}$ sends the transaction directly to $s$
and $t$, and the relevant routes are therefore of the form $a_\text{main} \to s \to u$ or
$a_\text{main} \to t \to u$; hence only $s$, $t$, and their direct neighbors may receive a new
transaction. Deliveries from $a_\text{redundancy}$ are not problematic because they are never
the first delivery of that transaction. The attacker therefore has $a_\text{redundancy}$ also
forward the transaction to any previously separated nodes among them, keeping their redundancy
levels above the target range and preventing route resets.

\subsubsection{Recognizing Active Routes}

Each per-route round assumes that the target route is active, meaning that it is the fastest
unfiltered route. The attacker may not know the current filtering state or which of several
routes with equal latency carries the transaction.

To determine when to proceed, the attacker repeats each round and uses an additional node
$a_\text{verification}$ directly connected to $t$. This node forwards the transaction from
$a_\text{main}$ to $t$ shortly after the expected delivery through $a_\text{main} \to s \to p
\to t$. If $s \to p \to t$ or a faster route remains active, its redundant delivery triggers
the \texttt{haveTx} message and causes that route to be filtered. If all such routes are
already filtered, $a_\text{verification}$ provides the first redundant forwarding delivery and
therefore receives the \texttt{haveTx} message, indicating that the attacker can proceed to the
next target route. This procedure does not require knowledge of the active routes or the exact
path taken by the transaction.

\subsubsection{Dealing with the Redundancy Control Interval}
\label{sec:dealing-with-redundancy-control-interval}

The deliveries from $a_\text{redundancy}$ must occur in the same redundancy control interval
as the initial delivery to keep the measured redundancy above the target. The attacker must
therefore monitor or influence the timing of the intervals.

In the current DOG implementation, after a node has been idle for several intervals, its next
message restarts the redundancy control phase. The attacker can therefore leave a node idle
before sending it a new transaction from $a_\text{main}$. That transaction begins a restarted
interval, giving the deliveries from $a_\text{redundancy}$ sufficient time to occur within the
same interval.

\subsubsection{Dealing with Triangles}
\label{sec:dealing-with-triangles}

The method requires additional handling when $s$, $p$, and $t$ form a triangle, so that the
direct connection $s - t$ exists alongside the route $s \to p \to t$. Because the attacker may
be unable to filter the direct route $s \to t$, the longer route $s \to p \to t$ may never
become active and therefore may also remain unfiltered.

To remove the effect of the direct connection during later rounds, the attacker temporarily
filters the route $a_\text{main} \to s \to t$. For this purpose, $a_\text{main}$ does not
reconnect while the attacker processes all routes $s \to * \to t$ for a fixed $s$.

This method requires the following form of the triangle inequality:
$$
  w(s, p) < d(s, t) + w(p, t), \quad \forall p \in \mathrm{N}(s) \cap \mathrm{N}(t)\text{,}
$$
where $\mathrm{N}(u)$ denotes the neighbors of node $u$, $w(u, v)$ the latency of the direct
link $u - v$, and $d(u, v)$ the latency of the fastest route from $u$ to $v$.

The inequality allows $s \to p \to t$ to remain active while $a_\text{main} \to s \to t$ is
filtered: a transaction from $a_\text{main}$ reaches $p$ through $s$ before reaching it through
$t$.

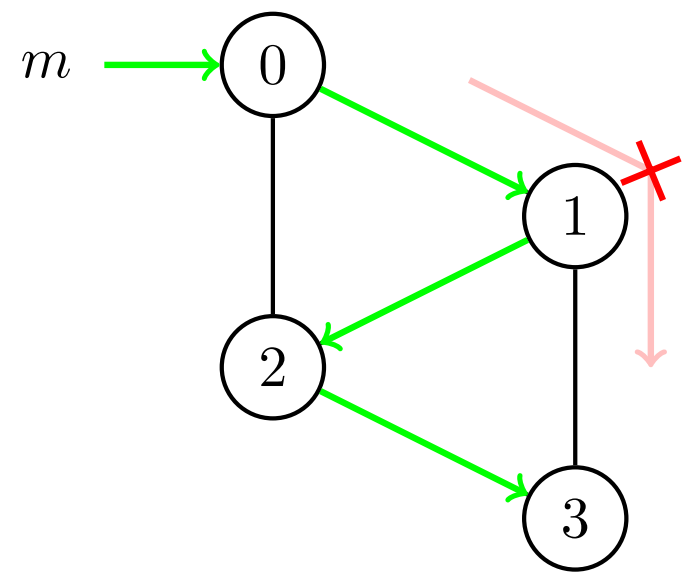
\begin{figure}[htb]
  \centering
  \input{figures/graphs/dealing-with-triangle.tex}
  \caption{An example of not filtering out routes in triangles.}
  \label{fig:dealing-with-triangles}
\end{figure}

Filtering $s \to p \to t$ remains necessary even when the direct route $s \to t$ cannot be
filtered, because otherwise transactions may still propagate beyond direct neighbors. For
example, Figure~\ref{fig:dealing-with-triangles} shows how node $0$ can reach node $3$, which
is not its direct neighbor, through an unfiltered route in a triangle.

\subsection{Verification}

We implemented the attack and evaluated it in a network simulated with
Shadow~\cite{shadow-web, shadow-ndss, shadow-atc}. The experiment codebase is available
at~\cite{li-code}. The simulated network consisted of 40 nodes, each of which established
outbound connections to 4 randomly selected nodes. After applying the attack to the full
network, all forwarding routes were filtered. Consequently, a transaction could still be
delivered directly from its publisher to the publisher's neighbors, but it was not forwarded
further by any receiving node.

\subsection{Result Summary and Cost}

The attack separates the full network, and the method in
Section~\ref{sec:maintaining-separation} maintains that state. Transactions may still travel
directly from their publishers to neighboring nodes because the attack may not filter direct
routes, but no receiving node forwards them further. DOG therefore no longer provides
multi-hop transaction dissemination. Together with the costs below, this demonstrates a
practical vulnerability in DOG under the stated assumptions.

Attacking a single route requires publishing $O(1)$ transactions and $O(\Delta)$ additional
deliveries by $a_\text{redundancy}$, where $\Delta$ is the maximum node degree. The attack
takes $O(T)$ time, where $T$ is the redundancy control interval. This matches the lower bound
because a target permits at most one \texttt{haveTx} message per interval.

Attacking the full network requires attacking each of its $R = O(n\Delta^2)$ routes, where $n$
is the number of nodes. The attack therefore publishes $O(R)$ transactions, requires
$O(R\Delta)$ additional deliveries by $a_\text{redundancy}$, and takes $O(RT)$ time when routes
are attacked sequentially.

Maintaining the separation requires forwarding each transaction published by a node $s$ to its
direct neighbors, at a cost of $O(\deg(s)) = O(\Delta)$ deliveries per transaction. Since
publishing a transaction already requires $\Omega(\deg(s))$ direct deliveries by $s$, the
asymptotic maintenance cost matches the publication cost.

\subsection{Condition Analysis and Extensions}

This section states the assumptions under which the attack operates and discusses possible
extensions that relax them.

\subsubsection{Basic Assumptions}

The attack relies on the following capabilities, which are natural for ordinary participants
in the P2P network:
\begin{itemize}
  \item The attacker controls $O(1)$ nodes connected to the network.
  \item These nodes can connect to any network node.
  \item These nodes can publish their own transactions.
  \item These nodes can delay transactions that they publish or forward.
\end{itemize}

The attacker must additionally know the network topology in order to identify the target
routes.

\subsubsection{Scenario Simplifications}

The analysis also makes the following simplifying assumptions, several of which can be relaxed
with additional implementation effort.
\begin{itemize}
  \item \emph{The network topology is static.} This assumption may be relaxed by applying the
        attack to newly created routes when the topology changes.
  \item \emph{The network latency is stable.} This assumption may be relaxed because the
        selected delays tolerate some latency variation; if a fluctuation prevents a round from
        succeeding, the attacker can also repeat that round.
  \item \emph{Other nodes remain silent during the attack.} This assumption may be relaxed by
        handling transactions published by other nodes with the maintenance method from
        Section~\ref{sec:maintaining-separation}. If such traffic restores a route, the attacker
        can also repeat the corresponding round.
  \item \emph{The attacker can monitor or control the redundancy control phases.} This
        assumption may be relaxed by inferring phase changes from \texttt{haveTx} messages.
        Section~\ref{sec:dealing-with-redundancy-control-interval} also gives a method specific
        to the current DOG implementation. If the timing is incorrect and a route is reset, the
        attacker can also repeat the round.
  \item \emph{For triangle links, the triangle inequality from
        Section~\ref{sec:dealing-with-triangles} holds.} This assumption allows routes within
        triangles to be filtered using the method described there. Other methods may relax it,
        for example by manipulating the filtering of routes $a_\text{main} \to t \to *$, but they
        are not analyzed here.
\end{itemize}

\subsubsection{Malicious Behaviors}

The implementation described above deviates from faithful DOG behavior in two ways, although
both can be replaced by actions available to faithful nodes. However, in either way, the
attacker does not need to create invalid transactions; it publishes ordinary transactions.
\begin{itemize}
  \item An attacker node may ignore \texttt{haveTx} messages so that it continues forwarding as
        required, particularly when acting as $a_\text{redundancy}$. A faithful alternative is to
        reconnect after receiving a \texttt{haveTx} message, thereby resetting the affected
        routes from $a_\text{main}$.
  \item An attacker node may selectively send transactions to particular nodes. A faithful
        alternative is to use separate nodes for different routes. Selective delivery is also
        externally indistinguishable from packet loss or unstable connectivity.
\end{itemize}

\subsubsection{Parallelization}

The attack described above processes routes sequentially and therefore takes $O(RT)$ time,
where $R$ is the number of routes and $T$ is the redundancy control interval. Although the
per-route attack already matches the $O(T)$ lower bound, attacks against sufficiently separated
target nodes may be executed in parallel with limited interference. The achievable parallelism
depends on the network topology and introduces additional interference cases, so it is not
analyzed further here.

\subsubsection{Analysis of Potential Fixes}

We did not identify a direct mitigation that can be added to the current DOG design without
changing its underlying route-filtering behavior. The attack relies on the following
properties:
\begin{itemize}
  \item A filtered route is not reset while its target's redundancy level remains above the
        target range.
  \item Routes to a separated node are not reset automatically.
  \item Route-filtering decisions use only local information about neighboring nodes.
        Incorporating global information would introduce non-trivial scalability costs.
  \item An attacker-controlled node can become a forwarder on a preserved route, whether route
        preservation depends on latency or random selection.
\end{itemize}

Under these conditions, the attacker can seek to make attacker-controlled nodes the only
preserved forwarders to a target and then disconnect them to cause separation.

The first two conditions can be relaxed if route resets occur only at a limited rate: the
attacker can repeat the corresponding rounds to restore the separation.

%% file: figures/graphs/route-filtering-1.tex
\begin{tikzpicture}[circle,thick,nodes={draw}]
  \node (s1) at (-1,0) {$s_1$};
  \node (s2) at (1,0) {$s_2$};
  \node (p1) at (-1,-1.5) {$p_1$};
  \node (u1) at (-1,-3) {$u_1$};
  \node (p2) at (1,-1.5) {$p_2$};
  \node (u2) at (1,-3) {$u_2$};
  \node (t) at (0,-4) {$t$};

  \draw (s1) -- (p1) -- (u1) -- (t);
  \draw (s2) -- (p2) -- (u2) -- (t);
  \draw (s1) -- (p2);
  \draw (s2) -- (p1);

  \begin{scope}[very thick, ->]
    \node[draw=none] (m1) at (-2.5,0) {$m_1$};

    \draw[green] (m1) -- (s1);

    \path[green]
    (s1) edge (p1)
    (p1) edge (u1)
    (u1) edge (t)
    (s1) edge (p2)
    (p2) edge (u2);

    \draw[blue, dashed]
    (u2.190) -- (t.90);

    \draw[blue]
    ([xshift=.5cm] $(p2)!0.75!(u2)$)
    to ([xshift=.5cm,yshift=-.5cm] u2) node[cross out,red,rotate=67.5] {}
    to ([xshift=.5cm,yshift=-.5cm] $(u2)!0.5!(t)$);
  \end{scope}
\end{tikzpicture}

%% file: figures/graphs/route-filtering-2.tex
\begin{tikzpicture}[circle,thick,nodes={draw}]
  \node (s1) at (-1,0) {$s_1$};
  \node (s2) at (1,0) {$s_2$};
  \node (p1) at (-1,-1.5) {$p_1$};
  \node (u1) at (-1,-3) {$u_1$};
  \node (p2) at (1,-1.5) {$p_2$};
  \node (u2) at (1,-3) {$u_2$};
  \node (t) at (0,-4) {$t$};

  \draw (s1) -- (p1) -- (u1) -- (t);
  \draw (s2) -- (p2) -- (u2) -- (t);
  \draw (s1) -- (p2);
  \draw (s2) -- (p1);

  \begin{scope}[very thick, ->]
    \node[draw=none] (m2) at (2.5,0) {$m_2$};

    \draw[green] (m2) -- (s2);

    \path[green]
    (s2) edge (p1)
    (p1) edge (u1)
    (u1) edge (t)
    (s2) edge (p2)
    (p2) edge (u2);

    \draw[pink]
    ([xshift=.5cm] $(p2)!0.75!(u2)$)
    to ([xshift=.5cm,yshift=-.5cm] u2) node[cross out,red,rotate=67.5] {}
    to ([xshift=.5cm,yshift=-.5cm] $(u2)!0.5!(t)$);
  \end{scope}
\end{tikzpicture}

%% file: figures/graphs/simple-ring-example-1.tex
\begin{tikzpicture}[circle,thick,nodes={draw}]
  \node (0) at (0,0) {$0$};
  \node (1) at (0,-4) {$1$};
  \node (2) at (0,-2) {$2$};
  \node (3) at (2,-1) {$3$};
  \node (4) at (2,-3) {$4$};

  \draw (0) -- (2) -- (1);
  \draw (0) -- (3) -- (4) -- (1);

  \begin{scope}[very thick, ->]
    \node[draw=none] (m1) at (-1.5,0) {$m_1$};

    \draw[green] (m1) -- (0);

    \path[green]
    (0) edge (2)
    (2) edge (1)
    (0) edge (3)
    (3) edge (4);

    \draw[blue, <->, dashed]
    (1.60) -- (4.180);

    \draw[blue]
    ([xshift=.4cm] $(3)!0.75!(4)$)
    to ([xshift=.4cm,yshift=-.3cm] 4) node[cross out,red,rotate=100] {}
    to ([xshift=.4cm,yshift=-.3cm] $(4)!0.4!(1)$);

    \draw[blue]
    ([xshift=-.4cm] $(2)!0.75!(1)$)
    to ([xshift=-.4cm,yshift=-.6cm] 1) node[cross out,red,rotate=22.5] {}
    to ([xshift=-.4cm,yshift=-.6cm] $(1)!0.5!(4)$);
  \end{scope}
\end{tikzpicture}

%% file: figures/graphs/simple-ring-example-2.tex
\begin{tikzpicture}[circle,thick,nodes={draw}]
  \node (0) at (0,0) {$0$};
  \node (1) at (0,-4) {$1$};
  \node (2) at (0,-2) {$2$};
  \node (3) at (2,-1) {$3$};
  \node (4) at (2,-3) {$4$};

  \draw (0) -- (2) -- (1);
  \draw (0) -- (3) -- (4) -- (1);

  \begin{scope}[very thick, ->]
    \draw[pink]
    ([xshift=.4cm] $(3)!0.75!(4)$)
    to ([xshift=.4cm,yshift=-.3cm] 4) node[cross out,red,rotate=100] {}
    to ([xshift=.4cm,yshift=-.3cm] $(4)!0.4!(1)$);

    \draw[pink]
    ([xshift=-.4cm] $(2)!0.75!(1)$)
    to ([xshift=-.4cm,yshift=-.6cm] 1) node[cross out,red,rotate=22.5] {}
    to ([xshift=-.4cm,yshift=-.6cm] $(1)!0.5!(4)$);

    \node[draw=none] (m2) at (-1.5,-4) {$m_2$};

    \draw[green] (m2) -- (1);

    \path[green]
    (1) edge (2)
    (2) edge (0)
    (1) edge (4)
    (4) edge (3);

    \draw[blue, <->, dashed]
    (0.-60) -- (3.180);

    \draw[blue]
    ([xshift=.4cm] $(4)!0.75!(3)$)
    to ([xshift=.4cm,yshift=.3cm] 3) node[cross out,red,rotate=-100] {}
    to ([xshift=.4cm,yshift=.3cm] $(3)!0.4!(0)$);

    \draw[blue]
    ([xshift=-.4cm] $(2)!0.75!(0)$)
    to ([xshift=-.4cm,yshift=.6cm] 0) node[cross out,red,rotate=-22.5] {}
    to ([xshift=-.4cm,yshift=.6cm] $(0)!0.5!(3)$);
  \end{scope}
\end{tikzpicture}

%% file: figures/graphs/simple-ring-example-3.tex
\begin{tikzpicture}[circle,thick,nodes={draw}]
  \node (0) at (0,0) {$0$};
  \node (1) at (0,-4) {$1$};
  \node (2) at (0,-2) {$2$};
  \node (3) at (2,-1) {$3$};
  \node (4) at (2,-3) {$4$};

  \draw (0) -- (2) -- (1);
  \draw (0) -- (3) -- (4) -- (1);

  \begin{scope}[very thick, ->]
    \draw[pink]
    ([xshift=.4cm] $(3)!0.75!(4)$)
    to ([xshift=.4cm,yshift=-.3cm] 4) node[cross out,red,rotate=100] {}
    to ([xshift=.4cm,yshift=-.3cm] $(4)!0.4!(1)$);

    \draw[pink]
    ([xshift=-.4cm] $(2)!0.75!(1)$)
    to ([xshift=-.4cm,yshift=-.6cm] 1) node[cross out,red,rotate=22.5] {}
    to ([xshift=-.4cm,yshift=-.6cm] $(1)!0.5!(4)$);

    \draw[pink]
    ([xshift=.4cm] $(4)!0.75!(3)$)
    to ([xshift=.4cm,yshift=.3cm] 3) node[cross out,red,rotate=-100] {}
    to ([xshift=.4cm,yshift=.3cm] $(3)!0.4!(0)$);

    \draw[pink]
    ([xshift=-.4cm] $(2)!0.75!(0)$)
    to ([xshift=-.4cm,yshift=.6cm] 0) node[cross out,red,rotate=-22.5] {}
    to ([xshift=-.4cm,yshift=.6cm] $(0)!0.5!(3)$);

    \node[draw=none] (m3) at (-1.5,-2) {$m_3$};

    \draw[green] (m3) -- (2);

    \path[green]
    (2) edge (0)
    (2) edge (1);

    \node[draw=none,red] at (1.5, -1) {?};
    \node[draw=none,red] at (1.5, -3) {?};

  \end{scope}
\end{tikzpicture}

%% file: figures/graphs/generalized-redundancy-example.tex
\begin{tikzpicture}[circle,thick,nodes={draw}]
  \node (0) at (1,0) {$0$};
  \node (1) at (1,-6) {$1$};
  \node (2) at (1,-1.5) {$2$};
  \node (3) at (1,-3) {$3$};
  \node (4) at (1,-4.5) {$4$};
  \node (5) at (3,-1.5) {$5$};
  \node (6) at (3,-3) {$6$};
  \node (7) at (3,-4.5) {$7$};
  \node (8) at (5,-1.5) {$8$};
  \node (9) at (5,-3) {$9$};
  \node (10) at (5,-4.5) {$10$};

  \draw (0) -- (2) -- (3) -- (4) -- (1);
  \draw (0) -- (5) -- (6) -- (7) -- (1);
  \draw (0) -- (8) -- (9) -- (10) -- (1);
  \draw[very thick]
  (0) to[out=180,in=180]
  node[draw=none,midway,xshift=-1cm] {(fastest)} (1);

  \begin{scope}[very thick, ->]
    \draw[pink]
    ([xshift=-.5cm] $(3)!0.5!(4)$)
    to ([xshift=-.5cm] 4) node[cross out,red] {}
    to ([xshift=-.5cm] $(4)!0.5!(1)$);

    \draw[pink]
    ([xshift=-.5cm] $(3)!0.5!(2)$)
    to ([xshift=-.5cm] 2) node[cross out,red] {}
    to ([xshift=-.5cm] $(2)!0.5!(0)$);

    \draw[pink]
    ([xshift=-.5cm] $(6)!0.5!(7)$)
    to ([xshift=-.5cm] 7) node[cross out,red,rotate=22.5] {}
    to ([xshift=-.5cm] $(7)!0.4!(1)$);

    \draw[pink]
    ([xshift=.5cm] $(9)!0.5!(8)$)
    to ([xshift=.5cm,yshift=.3cm] 8) node[cross out,red] {}
    to ([xshift=.5cm,yshift=.3cm] $(8)!0.3!(0)$);

    \node[draw=none] (m) at (-0.5,-3) {$m$};

    \draw[green] (m) -- (3);

    \path[green]
    (3) edge (2)
    (3) edge (4);

    \node[draw=none] (m0) at (-1,0) {$m^{(0)} \cdots$};

    \draw[green!30!gray] (m0) -- (0);

    \node[draw=none] (m1) at (-1,-6) {$m^{(1)} \cdots$};

    \draw[green!30!gray] (m1) -- (1);

    \node[draw=none,red] at (1.2, -.75) {?};
    \node[draw=none,red] at (1.2, -5.25) {?};
  \end{scope}

\end{tikzpicture}

%% file: figures/graphs/potential-attack-example.tex
\begin{tikzpicture}[circle,thick,nodes={draw}]
  \node (0) at (0,0) {$0$};
  \node (1) at (0,-4) {$1$};
  \node[rectangle] (2) at (-4,-2) {$2$};
  \node[rectangle] (3) at (-2,-2) {$3$};
  \node (4) at (0,-2) {$4$};
  \node (5) at (2,-2) {$5$};
  \node (6) at (4,-2) {$6$};

  \draw (0) -- (2) -- (1);
  \draw (0) -- (3) -- (1);
  \draw (0) -- (4) -- (1);
  \draw (0) -- (5) -- (1);
  \draw (0) -- (6) -- (1);

  \begin{scope}[very thick, ->]
    \draw[pink]
    ([xshift=-.5cm] $(0)!0.5!(4)$)
    to ([xshift=-.5cm] 4) node[cross out,red] {}
    to ([xshift=-.5cm] $(4)!0.5!(1)$);

    \draw[pink]
    ([xshift=-.5cm] $(0)!0.6!(5)$)
    to ([xshift=-.5cm] 5) node[cross out,red,rotate=30] {}
    to ([xshift=-.5cm] $(5)!0.4!(1)$);

    \draw[pink]
    ([xshift=.8cm] $(0)!0.7!(6)$)
    to ([xshift=.8cm] 6) node[cross out,red,rotate=30] {}
    to ([xshift=.8cm] $(6)!0.3!(1)$);

    \node[draw=none] (m0) at (-2,0) {$m^{(0)} \cdots$};
    \draw[green!30!gray] (m0) -- (0);

    \node[draw=none] (m) at (1.5,0) {$m$};
    \draw[green] (m) -- (0);

    \path[green]
    (0) edge (4)
    (0) edge (5)
    (0) edge (6);

    \node[red, cross out] at (-4,-2.25) {};
    \node[red, cross out] at (-2,-2.25) {};

    \node[draw=none] at (-5.25,-2) {(fastest)};

    \node[draw=none,red] at (-.5,-4) {?};
  \end{scope}

\end{tikzpicture}

%% file: figures/graphs/attacking-single-route.tex
\begin{tikzpicture}[thick,nodes={draw}]
  \node[circle] (s) at (0,0) {$s$};
  \node[circle] (p) at (0,-2) {$p$};
  \node[circle] (t) at (0,-4) {$t$};

  \node (a_main) at (2,-2) {$a_\text{main}$};
  \node (a_redundancy) at (3,-4) {$a_\text{redundancy}$};

  \draw (s) -- (p) -- (t);

  \path
  (a_main) edge (s)
  (a_main) edge (t)
  (a_main) edge (a_redundancy)
  (a_redundancy) edge (t);

  \begin{scope}[very thick, ->]
    \node[draw=none] (m) at (4,-2) {$m$};
    \draw[green] (m) -- (a_main);

    \path[green]
    (a_main) edge (s)
    (a_main) edge (t)
    (a_main) edge (a_redundancy)
    (s) edge (p);

    \draw[blue, dashed] (p.-110) -- (t.110);
    \draw[blue, dotted] (a_redundancy.187.5) -- (t.-20);

    \draw[blue]
    ([xshift=-.5cm] $(s)!0.5!(p)$)
    to ([xshift=-.5cm] p) node[cross out,red] {}
    to ([xshift=-.5cm] $(p)!0.5!(t)$);

    \draw[orange,draw=none] (s) -- node[xshift=.25cm,draw=none,midway] {$w_1$} (p);
    \draw[orange,draw=none] (p) -- node[xshift=.25cm,draw=none,midway] {$w_2$} (t);
    \draw[orange,draw=none] (a_main) -- node[xshift=.5cm,draw=none,midway] {$L_s \ge l_s$} (s);
    \draw[orange,draw=none] (a_main) -- node[xshift=.5cm,draw=none,midway] {$L_t \ge l_t$} (t);
  \end{scope}

\end{tikzpicture}

%% file: figures/graphs/maintaining-separation.tex
\begin{tikzpicture}[circle,thick,nodes={draw}]
  \node (0) at (0,0) {$0$};
  \node (1) at (0,-2) {$1$};
  \node (2) at (0,-4) {$2$};

  \node[rectangle] (a_redundancy) at (3,-2) {$a_\text{redundancy}$};

  \draw (0) -- (1) -- (2);
  \draw (0) -- (a_redundancy);
  \draw (1) -- (a_redundancy);
  \draw (2) -- (a_redundancy);

  \begin{scope}[very thick, ->]
    \draw[<->,pink]
    ([xshift=1.5cm] $(0)!0.5!(a_redundancy)$)
    to ([xshift=1.5cm] a_redundancy) node[cross out,red,rotate=30] {}
    to ([xshift=1.5cm] $(a_redundancy)!0.5!(2)$);

    \node[draw=none] (m1) at (-1.5,0) {$m_1$};
    \draw[green] (m1) -- (0);

    \path[green]
    (0) edge (1)
    (0) edge (a_redundancy);

    \draw[green, dashed]
    (a_redundancy.170) -- (1.30);

    \node[draw=none] (m2) at (-1.5,-4) {$m_2$};
    \draw[cyan] (m2) -- (2);

    \path[cyan]
    (2) edge (1)
    (2) edge (a_redundancy);

    \draw[cyan, dashed]
    (a_redundancy.190) -- (1.-30);
  \end{scope}
\end{tikzpicture}

%% file: figures/graphs/dealing-with-triangle.tex
\begin{tikzpicture}[circle,thick,nodes={draw}]
  \node (0) at (0,0) {$0$};
  \node (1) at (2,-1) {$1$};
  \node (2) at (0,-2) {$2$};
  \node (3) at (2,-3) {$3$};

  \draw (0) -- (1) -- (2) -- (3);
  \draw (0) -- (2);
  \draw (1) -- (3);

  \begin{scope}[very thick, ->]
    \node[draw=none] (m) at (-1.5,0) {$m$};
    \draw[green] (m) -- (0);

    \path[green]
    (0) edge (1)
    (1) edge (2)
    (2) edge (3);

    \draw[pink]
    ([xshift=.5cm, yshift=.3cm] $(0)!0.4!(1)$)
    to ([xshift=.5cm, yshift=.3cm] 1) node[cross out,red,rotate=67.5] {}
    to ([xshift=.5cm] $(1)!0.5!(3)$);
  \end{scope}
\end{tikzpicture}

%% file: sections/bibliography.tex

%% file: main.bbl
\begin{thebibliography}{99}

\bibitem{informal-dog}
  Informal Systems, ``CometBFT DOG protocol reduces mempool bandwidth by 2.5x,''
  \url{https://informal.systems/blog/a-new-mempool-gossip-protocol-for-cometbft}, Dec.\ 2024,
  accessed: 2026-07-30.

\bibitem{cometbft-dog-spec}
  CometBFT, ``Dynamic optimal graph (DOG) gossip protocol,''
  \url{https://github.com/cometbft/cometbft/blob/c451150275612dc4182e4d47f6722f2e49209829/spec/mempool/gossip/dog.md},
  accessed: 2026-07-30.

\bibitem{cometbft}
  CometBFT, ``CometBFT: A distributed, byzantine fault-tolerant, deterministic state machine
  replication engine. A fork and successor to tendermint core,''
  \url{https://github.com/cometbft/cometbft}, accessed: 2026-07-30.

\bibitem{gossipsub-spec}
  libp2p, ``gossipsub: An extensible baseline pubsub protocol,''
  \url{https://github.com/libp2p/specs/tree/master/pubsub/gossipsub}, accessed: 2026-07-30.

\bibitem{shadow-web}
  The Shadow team, ``The Shadow network simulator,'' \url{https://shadow.github.io}, accessed:
  2026-07-30.

\bibitem{shadow-ndss}
  R.\ Jansen and N.\ Hopper, ``Shadow: Running Tor in a box for accurate and efficient
  experimentation,'' in \emph{19th Annual Network and Distributed System Security Symposium
  (NDSS 2012)}, 2012.

\bibitem{shadow-atc}
  R.\ Jansen, J.\ Newsome, and R.\ Wails, ``Co-opting linux processes for high-performance
  network simulation,'' in \emph{2022 USENIX Annual Technical Conference (USENIX ATC 22)}.
  Carlsbad, CA: USENIX Association, Jul.\ 2022, pp.\ 327--350.

\bibitem{li-code}
  J.\ Li, ``Attacking p2p networks,'' \url{https://gitlab.ethz.ch/jiahli/attacking-p2p-networks}.

\bibitem{prysm}
  OffchainLabs, ``Prysm: An Ethereum consensus implementation written in Go,''
  \url{https://github.com/offchainlabs/prysm}, accessed: 2026-07-30.

\bibitem{gossipsub-v11}
  libp2p, ``gossipsub v1.1: Security extensions to improve on attack resilience and
  bootstrapping,''
  \url{https://github.com/libp2p/specs/blob/master/pubsub/gossipsub/gossipsub-v1.1.md},
  accessed: 2026-07-30.

\bibitem{doublezero}
  DoubleZero, ``The high-performance global network | DoubleZero,'' \url{https://doublezero.xyz},
  accessed: 2026-07-30.

\end{thebibliography}
